\PassOptionsToPackage{unicode}{hyperref}
\PassOptionsToPackage{hyphens}{url}
\PassOptionsToPackage{dvipsnames,svgnames,x11names}{xcolor}
\documentclass[12pt]{article}

\usepackage{amsmath,amssymb}
\usepackage{iftex}
\ifPDFTeX
  \usepackage[T1]{fontenc}
  \usepackage[utf8]{inputenc}
  \usepackage{textcomp} 
\else 
  \usepackage{unicode-math}
  \defaultfontfeatures{Scale=MatchLowercase}
  \defaultfontfeatures[\rmfamily]{Ligatures=TeX,Scale=1}
\fi
\usepackage{lmodern}
\ifPDFTeX\else  
\fi
\IfFileExists{upquote.sty}{\usepackage{upquote}}{}
\IfFileExists{microtype.sty}{
  \usepackage[]{microtype}
  \UseMicrotypeSet[protrusion]{basicmath} 
}{}
\makeatletter
\@ifundefined{KOMAClassName}{
  \IfFileExists{parskip.sty}{%
    \usepackage{parskip}
  }{
    \setlength{\parindent}{0pt}
    \setlength{\parskip}{6pt plus 2pt minus 1pt}}
}{
  \KOMAoptions{parskip=half}}
\makeatother
\usepackage{xcolor}
\makeatletter
\ifx\paragraph\undefined\else
  \let\oldparagraph\paragraph
  \renewcommand{\paragraph}{
    \@ifstar
      \xxxParagraphStar
      \xxxParagraphNoStar
  }
  \newcommand{\xxxParagraphStar}[1]{\oldparagraph*{#1}\mbox{}}
  \newcommand{\xxxParagraphNoStar}[1]{\oldparagraph{#1}\mbox{}}
\fi
\ifx\subparagraph\undefined\else
  \let\oldsubparagraph\subparagraph
  \renewcommand{\subparagraph}{
    \@ifstar
      \xxxSubParagraphStar
      \xxxSubParagraphNoStar
  }
  \newcommand{\xxxSubParagraphStar}[1]{\oldsubparagraph*{#1}\mbox{}}
  \newcommand{\xxxSubParagraphNoStar}[1]{\oldsubparagraph{#1}\mbox{}}
\fi
\makeatother

\usepackage{longtable,booktabs,array}
\usepackage{calc} 
\usepackage{etoolbox}
\makeatletter
\patchcmd\longtable{\par}{\if@noskipsec\mbox{}\fi\par}{}{}
\makeatother
\IfFileExists{footnotehyper.sty}{\usepackage{footnotehyper}}{\usepackage{footnote}}
\makesavenoteenv{longtable}
\usepackage{graphicx}
\makeatletter
\def\maxwidth{\ifdim\Gin@nat@width>\linewidth\linewidth\else\Gin@nat@width\fi}
\def\maxheight{\ifdim\Gin@nat@height>\textheight\textheight\else\Gin@nat@height\fi}
\makeatother
\setkeys{Gin}{width=\maxwidth,height=\maxheight,keepaspectratio}
\makeatletter
\def\fps@figure{htbp}
\makeatother

\makeatletter
\@ifpackageloaded{caption}{}{\usepackage{caption}}
\AtBeginDocument{%
\ifdefined\contentsname
  \renewcommand*\contentsname{Table of contents}
\else
  \newcommand\contentsname{Table of contents}
\fi
\ifdefined\listfigurename
  \renewcommand*\listfigurename{List of Figures}
\else
  \newcommand\listfigurename{List of Figures}
\fi
\ifdefined\listtablename
  \renewcommand*\listtablename{List of Tables}
\else
  \newcommand\listtablename{List of Tables}
\fi
\ifdefined\figurename
  \renewcommand*\figurename{Figure}
\else
  \newcommand\figurename{Figure}
\fi
\ifdefined\tablename
  \renewcommand*\tablename{Table}
\else
  \newcommand\tablename{Table}
\fi
}
\@ifpackageloaded{float}{}{\usepackage{float}}
\floatstyle{ruled}
\@ifundefined{c@chapter}{\newfloat{codelisting}{h}{lop}}{\newfloat{codelisting}{h}{lop}[chapter]}
\floatname{codelisting}{Listing}

\makeatother
\makeatletter
\@ifpackageloaded{caption}{}{\usepackage{caption}}
\@ifpackageloaded{subcaption}{}{\usepackage{subcaption}}
\makeatother

\ifLuaTeX
  \usepackage{selnolig}  
\fi
\usepackage[]{natbib}
\usepackage{bookmark}
\usepackage{shortcuts}
\usepackage{tikz}
\usetikzlibrary{shapes.geometric} 

\IfFileExists{xurl.sty}{\usepackage{xurl}}{} 
\hypersetup{
  pdftitle={Estimation of Directed Acyclic Graphs by Frequentist Model Averaging},
  pdfauthor={Huihang Liu, Wenhui Li, and Xinyu Zhang},
  pdfkeywords={Network data, asymptotic optimality, model misspecification},
  colorlinks=true,
  linkcolor={blue},
  filecolor={Maroon},
  citecolor={Blue},
  urlcolor={Blue},
  pdfcreator={LaTeX via pandoc}}

\newcommand{\anon}{1}

\begin{document}

\def\spacingset#1{\renewcommand{\baselinestretch}%
{#1}\small\normalsize} \spacingset{1}
\date{}


\if1\anon
{
  \title{\bf Estimation of Directed Acyclic Graphs by Frequentist Model Averaging}
  \author{Huihang Liu$^a$, Wenhui Li$^{b,c}$, and Xinyu Zhang$^{b,c}$ \hspace{.2cm}\\[0.4cm]
    \parbox{\linewidth}{%
      \centering
      $^a$Institute of Big Data Research, School of Statistics and Data Science, Shanghai University of Finance and Economics\\
      $^{b}$Center for Forecasting Science, Academy of Mathematics and Systems Science, Chinese Academy of Sciences\\
      $^{c}$State Key Laboratory of Mathematical Sciences, Academy of Mathematics and Systems Science, Chinese Academy of Sciences
    }}
  \maketitle
} \fi

\if0\anon
{
  \bigskip
  \bigskip
  \bigskip
  \begin{center}
    {\LARGE\bf Estimation of Directed Acyclic Graphs by Frequentist Model Averaging}
\end{center}
  \medskip
} \fi

\bigskip
\begin{abstract}
 Directed acyclic graphs provide a fundamental tool for representing directed dependence structures in multivariate network data, and are widely used to model financial and economic networks. However, accurate and interpretable estimation remains challenging under graph structural uncertainty. We propose an optimal model averaging method for directed acyclic Gaussian graphs. With a set of candidate models varying by graph structures, we average estimates from candidate models using weights that minimize a penalized negative log-likelihood criterion. In contrast to existing approaches, we not only establish the asymptotic optimality, weight consistency, and parameter consistency of the proposed method, but also explicitly characterize how different candidate models affect the convergence rate. Moreover, we prove parameter consistency even when all candidate graph models are misspecified. Results from simulation studies and a real-data analysis on the banks' international liability data show the promise of the proposed method.
\end{abstract}

\noindent%
{\it Keywords:} asymptotic optimality, model misspecification, structure uncertainty
\vfill

\newpage
\spacingset{1.8} 
\section{Introduction} \label{introduction} 


The directed and acyclic structure is crucial in scenarios where causality and directional movement of variables are of interest, without the effects of loops that can obfuscate the true direction of dependencies. Such causal graphical diagrams are widely used in empirical economics \citep{card1999chapter,heckman2010building,gruber2016public,imbens2020Potential}. The Gaussian directed acyclic graph (DAG) model learned from intervention data is a standard tool in statistics and genomics \citep{ellis2008jasa,vanderweele2010jrssb,Han2016JASA}. DAG employs directed edges to represent the linkages among interacting entities and is designed to be loop-free, thereby preserving the acyclic nature of the network. Despite these successes, the underlying graph structure is typically unknown and must be learned from data.
Different plausible graphs can lead to different scientific conclusions, making results sensitive to graph selection and model uncertainty
\citep{leamer1983lets,durlauf2005growth}.

Several model selection methods based on DAG model have been proposed in recent years, including those leveraging regularization methods, as exemplified by \citet{yuan2019constrained}. These approaches operate under the premise that among the available candidate models, there exists a singular ``best'' or most accurate DAG model, and their objective is to identify this optimal model from the set. However, there is a risk that all the candidate models considered might be misspecified. This means that graph structures of candidate models do not accurately reflect the true underlying structure in the data generating process. Consequently, the choice of these misspecified models potentially leads to invalid subsequent analyzes. Alternatively, when multiple competitive models appear to fit the data with the same or similar accuracy, the question of which model to adopt becomes a subject of debate and controversy.

To address these challenges, we introduce an optimal model averaging estimation method for DAGs. 
With a variety of candidate models, each of which is distinguished by graph structures, we combine them to potentially provide a more accurate estimation of the directed and acyclic network.
This method acknowledges the limitations of individual models and mitigates the risk of relying on a single and potentially misspecified model.
The principle of model averaging has demonstrated success in various literature, as exemplified by \cite{Liang2011JASA}, \cite{hansen2014QE}, \cite{cheng2019QE}, and \cite{Lehrer2022MS}. However, these existing methods cannot be directly applied to DAGs due to DAGs' unique structural characteristics. 
Our approach specifically caters to these nuances, offering a tailored solution to the estimation challenges in DAG-based networks.

Our proposed method comprises three main steps. 
In Step 1, a set of nested candidate models is generated by the greedy search algorithm, and the graph structure of each candidate model is different. 
In Step 2, for each candidate model, we employ a structural equation model to learn the corresponding DAG. The coefficient matrix in the structural equation model can equivalently describe the node connection in the DAG \citep{shojaie2010Penalized,pearl2012causal,fu2013Learning,nagarajan2013bayesian,kaplan2016Bayesian,yuan2019constrained,suter2021BiDAG}. For each candidate DAG model, the coefficient matrix is estimated by regressing the targeted nodes on their parent nodes, motivated by \citet{meinshausen2006high}. Based on multiple estimates from the first two steps, in Step 3, we propose a penalized negative log-likelihood weight criterion for high-dimensional DAG models, which simultaneously solves the high-dimensional challenge and multivariate-response nature of DAG models.


Theoretically, we have four main contributions. 
First, unlike prior analyses on DAG that require a correctly specified graph to justify statistical guarantees, we demonstrate the parameter consistency of the DAG estimator in misspecified cases, where some true edges are missing from the estimated graphs. Second, we establish the asymptotic optimality of our proposed estimator in scenarios where all candidate models are misspecified, representing a key advancement in the model averaging literature for DAGs. Specifically, we show that the Kullback-Leibler loss of our estimator is asymptotically equivalent to that of the best but infeasible model averaging estimator.
Third, beyond high-level weight consistency statements commonly seen in earlier model averaging work, we provide the model-type-specific convergence rates for the estimated weights. Generally, we prove that the weights asymptotically concentrate on the minimal correctly specified graph, while the weights on underfitted and overfitted graphs vanish. Specifically, we characterize how these rates depend explicitly on the number of candidate graphs of each type, the edge complexity of the minimal correctly specified graph, the sample size, and a parsimonious penalty parameter.
Finally, we establish the convergence rate of the model-averaged parameter estimator by an error decomposition across correctly specified, underfitted, and overfitted candidates.

Numerically, the simulated experiments show the superiority of our proposed method over competing model selection and model averaging approaches across various scenarios. Empirically, we apply our method to model the network of national banks on a global scale, using data from the Bank for International Settlements. The resulting graph sheds light on the interdependencies and liabilities among the world's major economies. Our findings not only align with existing economic and financial literature, but also uncover new insights with significant economic implications.

The rest of this article is organized as follows:
Section~\ref{model-set-up-and-parametric-estimation} details the formulation of DAGs and the parameter estimation process. 
Section~\ref{model-averaging} introduces our proposed model averaging method. 
Section~\ref{theoretical-analysis} elaborates on the theoretical properties of this method. 
Section~\ref{numerical-study} presents numerical simulation results that demonstrate the performance of our proposed method and Section~\ref{real-data} applies the method to locational banking statistics for practical illustration. 
Section~\ref{sec:summary} summarizes this paper and presents some potential future works. 
Finally, the proofs of the related lemmas and theories of this article are presented in Appendix.

\section{Model Set-Up and Parameter Estimation} \label{model-set-up-and-parametric-estimation}

Suppose we have $p$ variables, denoted as $x_1,  \ldots, x_p$, and their joint probability distribution can be encoded by a DAG model, denoted as $\mc{G} = (V, E)$. In this representation, $V = \{1,  \ldots, p\}$ is the set of nodes, where each node corresponds to one random variable, and $E\subset \{ (k,j)  : k \neq j, k = 1,\ldots,p, \text{ and } j = 1,\ldots,p \}$ is the set of directed edges, which depicts the directional effects between nodes.  If  $(k, j) \in E$,  there is a directed edge with an arrow from $x_k$ pointing to $x_j$  in the graph, and we refer to the $k$\Th node as the parent node and the $j$\Th node as the child node. Denote the index set of parents of $x_j$ by $pa_j$ for $j = 1,\ldots,p$.

The directional relationships between variables induced by directed edges can be embedded into a linear structural equation model,
\begin{align}
  x_{j} = \sum_{k: k \neq j} x_{k} {A}_{k j} + z_{j}, \quad z_{j} \sim \mc{N}\left(0, \sigma^{2}\right), \quad \text{for } j\in\{1, \dots, p\},
  \label{eq:model}
\end{align}
where ${A}_{k j}$ corresponds to a directed edge from $x_k$ to $x_j$, $z_j$'s are independent Gaussian errors. 
When ${A}_{kj}\neq 0$, it signifies that the corresponding directed edge is included in $\mc{G}$, that is $(k,j)\in E$, and $x_k$ is a parent of $x_j$, that is, $k \in pa_j $. Conversely, when ${A}_{kj} = 0$, it means that $(k,j)\notin E$ and $k \notin pa_j $. 
Define the parameter matrix $\bA = \{{A}_{k j}\}_{1 \le k,j \le p}$. 
Our target is to estimate $\bA$ within the constraints of directional relations and acyclicity in a DAG model.

Given $n$ observations of each variable,  we denote the $i$\Th observation of $x_j$ and its corresponding random error by $x_{ij}$ and $z_{ij}$, respectively. Let $\bX = (\bx_1,\ldots,\bx_p)$ with $\bx_j = (x_{1j},\ldots, x_{nj})^{\top}$, and $\bZ = (\bz_1,\ldots,\bz_p)$ with $\bz_j = (z_{1j},\ldots, z_{nj})^{\top}$. Based on Model \eqref{eq:model}, we have 
\begin{align}
  \label{eq:modeleq}
  \bx_{j} = \sum_{k: k \neq j} \bx_{k} A_{kj} + \bz_{j},  \quad \text{for } j\in\{1, \dots, p\}.
\end{align}
Model \eqref{eq:modeleq} was also imposed in \citet{yuan2019constrained} and \citet{li2020likelihood} for DAG recovery. The parameter $\bA$ is estimated by the following log-likelihood function,
\begin{align}\label{eq:log-loss}
 \ell_n (\bA)   =  -\frac{n p}{2} \log \left(2 \pi \sigma^{2}\right)-\frac{n}{2 \sigma^{2}} \tr\left\{(\bI-\bA) (\bI-\bA)\t \hbSig_{n}\right\} ,
\end{align}
where $\hbSig_n = \bX\t\bX / n$ is the sample covariance matrix of $\bX$. By dropping some constant terms, the $-2 \times \ell_n (\bA)$ is left with
\begin{align} 
    n \tr \left\{(\bI-\bA) (\bI-\bA)\t \hbSig_n \right\}  
    = \sum_{j=1}^{p} \sum_{i=1}^{n}\left(x_{i j}-\sum_{k=1}^p x_{i k} A_{kj}\right)^{2} 
    = \| \bX - \bX \bA \|_{\mathrm{F}}^2. 
  \label{eq:loss}
\end{align}

\begin{remark} 
 The identifiability of DAGs has been studied under various assumptions about error variances. \citet{peters2014identifiability} demonstrated that a DAG can be identified if the Gaussian errors have equal variances. Later, \citet{ghoshal2018learning} and \citet{park2020identifiability} extended this result to cases involving unequal error variances, allowing for Gaussian and non-Gaussian noise, respectively, but they still required an explicit ordering of the noise variances. We focus on addressing the uncertainty in graph structure. To facilitate our theoretical analysis and ensure parameter identifiability, we adopt the assumption of equal variances for Gaussian errors. 
\end{remark}

Based on \eqref{eq:loss}, we can obtain $\hbA_{pa_j,j} = (\bX_{pa_j}\t \bX_{pa_j})^{-1} \bX_{pa_j}\t \bx_{j}$ and $\hbA_{-pa_j,j} = \bzero$, where  $\bX_{pa_j}$ is the submatrix of $\bX$ with columns indexed in $pa_j$, $\hbA_{pa_j}$ and  $\hbA_{-pa_j}$ are the submatrices of $\hbA$ with rows indexed in $pa_j$ and in $\{1,\ldots,p\}\setminus pa_j $, respectively, and $\hbA_{pa_j,j}\t$ and $\hbA_{-pa_j,j}\t$ are the $j$\Th column vectors of  $\hbA_{pa_j}$ and $\hbA_{-pa_j}$, respectively.   Note that if $pa_j$ is an empty set, then $\bX_{pa_j}$ and $\widehat{\bA}_{pa_j}$ are empty. Let $\bPi_j$ be a selection matrix of the $j$\Th node, where the element of the $i$\Th row and the $k$\Th column equals 1 if $i$ is the $k$\Th element of $pa_j$, that is ${(\bPi_j)}_{i,k} = 1$ if $i=(pa_j)_k$, and ${(\bPi_j)}_{i,k} = 0$ otherwise. Then the estimator of the $j$\Th column of $\bA$ is
\begin{align}
\hbA_j 
= \bPi_j (\bX_{pa_j}\t \bX_{pa_j})^{-1} \bX_{pa_j}\t \bx_{j} .
\label{eq:Aj.hat}
\end{align}
With \eqref{eq:Aj.hat}, the estimator of $\bA$ is $\hbA=(\hbA_1, \dots, \hbA_p)$. Similar techniques have been used in \citet{meinshausen2006high}.  Let the projection matrix of the $j$\Th node's parents be $\bP_j = \bX_{pa_j}(\bX_{pa_j}\t \bX_{pa_j})^{-1} \bX_{pa_j}\t$, and then we have $\bX_{pa_j} \hbA_{pa_j,j} = \bP_j \bx_{j}$ and $\bX \hbA = (\bP_1 \bx_{1}, \dots, \bP_p \bx_{p})$. 

The above procedure is conducted based the known edge set $\bE$. However, in reality, the true edge set is always unknown and there exists a set of candidate graphs with varying graph structures, specified by edge sets. We address the model uncertainty of which pairs of nodes have directed edges by a model averaging procedure. In the next section, we will discuss the proposed method in details.

\section{Model Averaging Procedure}\label{model-averaging}
 
Suppose there are $M_n$ DAGs varying by different edge sets, where $M_n$ is allowed to diverge to with the sample size $n$. 
A special case arises when nodes are treated as units of model uncertainty; that is, all edges connected to a node are either included in the model or excluded from it. This strategy has been adopted by classic model averaging methods in regression problems \citep{hansen2007least,zhang2011focused,zhang2016optimal,zouOptimalModelAveraging2022a}. However, such an approach does not adequately capture the full complexity of graph structures. To overcome this limitation, we develop a new model averaging method that leverages ideas from greedy search algorithms \citep{heckerman1995learning,chickering2002optimal,tsamardinos2006maxmin} and generates candidate models through an adaptive perturbation procedure. The procedure unfolds in three main steps, which we describe below.

\textbf{Step 1}: Generate a set of candidate graphs with different edge sets. We use a greedy forward/backward method to prepare $M_n$ nested DAGs. We start by obtaining an initial graph $\wh{\mc{G}}^0$ using some model selection method, such as the approach proposed by \citet{yuan2019constrained}. Then, $M_n-1$ nested graphs are generated through a greedy forward/backward search centered around the pre-selected graph $\wh{\mc{G}}^0$. Specifically, the data is split into two equal-sized parts: a training set and a validation set. For a given edge set $E$, we estimate the coefficient matrix on the training set and assess its profile likelihood on the validation set. The estimate of $\bA$ under edge set $E$ is denoted as $\hbA(E)$. 
In the forward phase, $[(M_n-1)/2]$ edges are added to maximize the validation profile likelihood, while in the backward phase, edges with minimal impact on the likelihood are removed. This process yields $M_n$ nested candidate models, denoted by ${\wh{\mc{G}}^{(1)}, \dots, \wh{\mc{G}}^{(M_n)}}$, with edge sets satisfying $E^{(1)} \subset \dots \subset E^{(M_n)}$.

\textbf{Step 2}: Estimate $\bA$ for each candidate graph. Given the predetermined graph structures from Step 1, we calculate an estimate of $\bA$ for each model. For the $m$\Th model, the estimator of $\bA$ is defined as
\begin{align}\label{eq:A_m}
\hbA(E^{(m)})
\triangleq \hbA^{(m)}
\triangleq (\bPi_1^{(m)} \bP_1^{(m)}\bx_1, \dots, \bPi_p^{(m)} \bP_p^{(m)}\bx_p),
\end{align}
where $\bPi_j^{(m)}$ and $\bP_j^{(m)}$ are the selection matrix and the projection matrix for node $j$ and edge set $E^{(m)}$, respectively. Specifically, let $pa_j^{(m)}$ denote the set of parent nodes of the $j$\Th node in graph $\wh{\mc{G}}^{(m)}$, and let $k_m$ represent the cardinality of $E^{(m)}$. We define $\bX^{(m)}_{pa_j}$ as the submatrix of $\bX$ with columns indexed by $pa_j^{(m)}$. Thus, $\bP_j^{(m)} = \bX_{pa_j}^{(m)} [(\bX_{pa_j}^{(m)})\t \bX_{pa_j}^{(m)} ]^{-1} (\bX_{pa_j}^{(m)})\t$.

\textbf{Step 3}: Combine the estimates to obtain the weighted estimator of $\bA$. The $M_n$ estimates of $\bA$ are combined to yield a weighted estimator:
\begin{align*}
\hbA(\bw)
= \sum_{m=1}^{M_n} w_m \hbA^{(m)},
\end{align*}
where $w_m$ is the $m$\Th element of the weight vector $\bw = (w_1, \dots, w_{M_n})$, which belongs to the simplex $\mc{W} = \{\bw \in [0,1]^{M_n} : \sum_{m=1}^{M_n} w_m = 1\}$.

It is important to note that loops may be created during the combination process if directed edges from various candidate models are linked sequentially. Therefore, a nested candidate structure is employed in Step 1 to ensure the acyclic constraint in the combined estimate. Additionally, using nested models helps reduce computational complexity since the total number of candidate models is smaller compared to the number that would result from considering all possible graph structures. This nested approach has been previously applied in model averaging research \citep{hansen2007least,zhang2020parsimonious}.
The detailed process is presented in Algorithm \ref{alg:MA-DAG}.

\begin{algorithm} 
  \caption{Model Averaging Estimation for DAG} \label{alg:MA-DAG}
  \begin{algorithmic}[1]
  \REQUIRE Data matrix $\bX$, parameters $M_n$ and $\lambda_n$
  \ENSURE Model averaging estimator $\hbA(\hbw)$

  \STATE Get an initial estimator $\wh{\mc{G}}^0$ and set $\wh{\mc{G}}^{(\lceil M_n /2\rceil)}=\wh{\mc{G}}^0=(V, \wh{E}^{(\lceil M_n/2\rceil)})$
  \STATE Split the data into equal-sized training and validation sets, i.e., $\mc{D}_{t}$ and $\mc{D}_{v}$

  \FOR{$i=1$ to $\lceil (M_n-1)/2 \rceil$}
    \STATE On $\mc{D}_t$, calculate $\wt{\bA}_e \triangleq \hbA(\wh{E}^{(\lceil M_n/2 \rceil+i-1)} \cup \{e\})$ with an edge $e \notin \wh{E}^{(\lceil M_n/2 \rceil+i-1)}$ and calculate $\wt{\sigma}^2_e$
    \STATE Calculate \begin{align*}
      \wh{e}^{\lceil M_n/2 \rceil+i} = \argmax_{e \notin \wh{E}^{(\lceil M_n/2 \rceil+i-1)}}  \frac{n p}{4} \log \left(2 \pi \wt{\sigma}^2_e \right) + \frac{n}{4 \wt{\sigma}^2_e} \tr\left\{(\bI-\wt{\bA}_e) (\bI-\wt{\bA}_e)\t \wt{\bSig}\right\},
    \end{align*} 
    where $\wt{\bSig}$ is the sample covariance matrix on $\mc{D}_v$
    \STATE $\wh{E}^{(\lceil M_n/2 \rceil+i)} = \wh{E}^{(\lceil M_n/2 \rceil+i-1)} \cup \{\wh{e}^{\lceil M_n/2 \rceil+i}\}$  and $\wh{\mc{G}}^{(\lceil M_n/2 \rceil+i)} = (V, \wh{E}^{(\lceil M_n/2 \rceil+i)})$
  \ENDFOR

  \FOR{$i=1$ to $\lceil  M_n/2 - 1 \rceil$}
    \STATE  On $\mc{D}_t$, calculate $\wt{\bA}_e \triangleq \hbA(\wh{E}^{(\lceil M_n/2 \rceil-i+1)} \setminus \{e\})$ with an edge $e \in \wh{E}^{(\lceil M_n/2 \rceil-i+1)}$ and calculate   $\wt{\sigma}^2_e$
    \STATE Calculate \begin{align*}
      \wh{e}^{\lceil M_n/2 \rceil-i} = \argmax_{e \in \wh{E}^{(\lceil M_n/2 \rceil-i+1)}}  \frac{n p}{4} \log \left(2 \pi \wt{\sigma}^2_e \right) + \frac{n}{4 \wt{\sigma}^2_e} \tr\left\{(\bI-\wt{\bA}_e) (\bI-\wt{\bA}_e)\t \wt{\bSig}\right\},
    \end{align*} 
    where $\wt{\bSig}$ is the sample covariance matrix on $\mc{D}_v$
    \STATE $\wh{E}^{(\lceil M_n/2 \rceil-i)} = \wh{E}^{(\lceil M_n/2 \rceil-i+1)} \cup \{\wh{e}^{\lceil M_n/2 \rceil-i}\}$  and $\wh{\mc{G}}^{(\lceil M_n/2 \rceil-i)} = (V, \wh{E}^{(\lceil M_n/2 \rceil-i)})$
  \ENDFOR

  \STATE Calculate $\hbA^{(m)} = \hbA(\wh{E}^{(m)})$ for each $m\in[M_n]$ using $\mc{D}_t \cup \mc{D}_v$
  \STATE Compute the weight $\hbw = (\wh{w}_1,\dots,\wh{w}_{M_n})\t$ using \eqref{eq:weight}

  \RETURN $\hbA(\hbw) = \sum_{m=1}^{M_n} \hbA^{(m)} \wh{w}_m$
  \end{algorithmic}
\end{algorithm}

We propose a penalized negative log-likelihood weight criterion, which incorporates a penalty term $\widetilde{\lambda}_n \bw\t \bk$ into $-2\times \ell_n (\hbA(\bw))$ as detailed in \eqref{eq:log-loss}. Here, we set $\widetilde{\lambda}_n = \lambda_n / \sigma^2$, with $\lambda_n$ being a tuning parameter that varies with $n$. 
Specifically, the criterion is
\begin{align} 
  \begin{aligned}[b]
    \mc{P}_n(\bw) 
    & = -2 \ell_n (\hbA(\bw)) + \widetilde{\lambda}_n \bw\t \bk \\
    & = n p \log \left(2 \pi \sigma^{2}\right) + \frac{n}{\sigma^{2}} \tr\Bigl[\bigl\{\bI-\hbA (\bw)\bigr\} \bigl\{\bI-\hbA (\bw)\bigr\}\t \hbSig_{n}\Bigr] + \widetilde{\lambda}_n \bw\t \bk \\
    & = n p \log \left(2 \pi \sigma^{2}\right) + \frac{1}{\sigma^{2}} \bigl\{ \| \bX - \bX \hbA(\bw) \|_{\mathrm{F}}^2 + \lambda_n \bw\t \bk \bigr\},
  \end{aligned}
  \label{eq:cirterion_penalized-likelihood}
\end{align}
where $\bk = (k_1, \dots, k_{M_n})\t$ is the vector of numbers of edges in candidate models. 
Let $\mc{C}_{n}(\bw) =\| \bX - \bX \hbA(\bw) \|_{\mathrm{F}}^2 + \lambda_n \bw\t \bk $. 
It is easy to see that $\mc{C}_{n} (\bw)$ can be seen as the multivariate extension of parsimonious model averaging \citep{zhang2020parsimonious}. 
When ${\lambda}_n=2$, $\mc{C}_{n}$ corresponds to $ \| \bX - \bX \hbA(\bw) \|_{\mathrm{F}}^2 + 2 \sigma^2 \bw\t \bk$, which is the multivariate extension of Mallows model averaging \citep{hansen2007least}. 
In practice, the variance $\sigma^2$ is unknown and can be estimated in the ``largest'' candidate model \citep{hansen2007least,zhang2020parsimonious}. 
Specifically, we estimate $\sigma^2$ by $\widehat{\sigma}^2 = \| \bX - \bX \hbA^{(M_n)} \|_2^2 / \{(n- k_{M_n}) p\} $.

Based on the weight criterion \eqref{eq:cirterion_penalized-likelihood}, we obtain the estimated weights by
\begin{align}
    \label{eq:weight}
    \hbw = \argmin_{\bw \in \mc{W}} \mc{P}_n(\bw). 
  \end{align}
Then, the proposed model averaging estimator of $\bA$ is given by $\hbA(\widehat{\bw})$.

\section{Theoretical Analysis} \label{theoretical-analysis}

In this section, we explore the theory of our proposed model averaging method. First, we introduce some notations.  Let $\bA_0$ be the true value of $\bA$ in Model \eqref{eq:model} and define $\bSig_0 = \sigma^2 (\bI - \bA_0\t)^{-1} (\bI - \bA_0)^{-1}$.  For the $m$\Th candidate model, we define $\bA^{(m)}$ as the adjacency matrix with the structure under the edge set $E^{(m)}$. We define the Kullback-Leibler  divergence to measure the divergence of using $\bA^{(m)}$ to approximate the true parameter matrix $\bA_0$ by,
\begin{align*}
  \begin{aligned}[b]
    \KL^{(m)} 
    & = \mb{E}_{\bX^*} \{\log(f(\bX^* \mid \bA_0, \sigma)) - \log(f({\bX}^* \mid \bA^{(m)} , \sigma))\} \\
    & = \frac{n}{2}  \left[\log \left|\boldsymbol{\Sigma}_0^{-1}\right|-\log |{\boldsymbol{\Omega}}^{(m)}|-p + \tr\left\{{\boldsymbol{\Omega}}^{(m)} \boldsymbol{\Sigma}_0\right\}\right]   \\
    & = \frac{n}{2 \sigma^{2}} \tr\left\{(\bI-\bA^{(m)} ) (\bI-\bA^{(m)} )\t \bSig_{0}\right\} - \frac{np}{2},
  \end{aligned}
\end{align*}
where $f ( \cdot | \bA, \sigma)$ is the Gaussian density of  with mean zero and covariance $\bSig = \sigma^2 (\bI - \bA\t)^{-1} (\bI - \bA)^{-1}$, $\bX^*$ is another realization distributed from $f ( \cdot | \bA_0, \sigma)$ and independent of $\bX$, and ${\boldsymbol{\Omega}}^{(m)} = (\bI - \bA^{(m)})(\bI - (\bA^{(m)})\t)/\sigma^2$ is the inverse covariance matrix corresponding to $\bA^{(m)}$. Let $\bA_*^{(m)}$ be the parameter matrix which minimizes the Kullback-Leibler  divergence $\KL^{(m)}$,
and $\bA_*^{(m)}$ is called the quasi-true value of the $m$\Th candidate model.
Let $\bA_*(\bw) = \sum_{m=1}^{M_n} w_m \bA_*^{(m)}$,
$\KL_*(\bw) = {n} \tr\left\{(\bI-\bA_*(\bw)) (\bI-\bA_*(\bw))\t \bSig_{0}\right\} / {2 \sigma^{2}} - np/2$,
and $\xi_n = \inf_{\bw\in\mc{W}} \KL_*(\bw)$.  All limiting processes discussed in this article assume the limit as $n \rightarrow \infty$.

\subsection{Asymptotic Optimality of Model Averaging Estimator} \label{asymptotic-optimality-of-model-averaging-estimator}
To build the asymptotic optimality of the proposed model averaging estimator, we first present some definitions. We define those models containing all edges in the true model (Model \eqref{eq:model}) as correctly specified models, and those models who lack at least one edge of the true model as  misspecified models.  Note that the pair $(k,j)$ is ordered, which is different from the tuple $(j,k)$.  Next,  we list some technique conditions to prove the asymptotic optimality. 

\begin{condition}  \label{con:parameter}
  There exists some positive constant $C$, such that ${1}/{C} \le s_{\min} (\bI-\bA_0) \le s_{\max} (\bI-\bA_0) \le C $, where $s_{\min}(\cdot)$ and $s_{\max}(\cdot)$ refer to the minimal and maximal singular values, respectively. 
\end{condition}

Condition \eqref{con:parameter} restricts the singular values of  $\bI-\bA_0$ to be bounded by positive constants from above and below. This is set to control the volatility range of {the random matrix $\bX$}.  Specifically, if we  take expectation to Equation~\eqref{eq:modeleq}, we can get $(\bI_p - \bA_0)\tran \bSig_{\bX} (\bI_p - \bA_0) = \sigma^2 \bI_p$,
where {$\bSig_{\bX}$ is the covariance matrix of $\bX$}.
Then Condition~\ref{con:parameter} can derive that the eigenvalues of $\bSig_{\bX}$ are bounded from above and below by some positive constants. Based on the fact that the elements of {$\bX\bSig_{\bX}^{-1/2}$} are i.i.d. from $\mc{N}(0,1)$, and the deviation inequality for extreme value of the eigenvalues of the Gaussian random matrix in  \cite{johnson2001handbook}, 
we have the following for any  $t \geq 0$, 
\begin{align*}
   \mathbb{P}\left(\sqrt{n}-\sqrt{p}-t \leq s_{\min}(\bX\bSig_{\bX}^{-1/2}) \leq s_{\max }(\bX \bSig_{\bX}^{-1/2}) \leq \sqrt{n}+\sqrt{p}+t\right) \geq 1-2 e^{-t^{2 } / 2},
\end{align*} 
which leads to
$\| \bX \|_2  \le \| \bSig_{\bX}^{1/2} \|_2\ \|\bX \bSig_{\bX}^{-1/2}\|_2 = O_p (n^{1/2})$, $\fnorm{\bX}
   \le \| \bSig_{\bX}^{1/2}\|_2 \ \fnorm{\bX \bSig_{\bX}^{-1/2}}
   = O_p (n^{1/2} p^{1/2})$,
and $s_{\min } (\bX) = O_p (n^{1/2})$.

\begin{lemma}[Consistency of candidate estimators] \label{lemma:candidate_consistency}
  If Condition~\ref{con:parameter} holds, then
  \begin{align*}
    \max_{1 \le m \le M_n} \| \hbA^{(m)} - \bA^{(m)}_* \|_\mathrm{F} = O_p (M_n^{1/2} k_{M_n}^{1/2} n^{-1/2}), 
  \end{align*}
  where $\hbA^{(m)}$ is calculated by Equation \eqref{eq:A_m}.  
\end{lemma}

Lemma \ref{lemma:candidate_consistency} shows that {the $m$\Th estimator of the parameter matrix} converges to the quasi-true value $\bA^{(m)}_*$ with the convergence rate of $M_n^{1/2} k_{M_n}^{1/2} n^{-1/2}$. 
We note that $k_{M_n}$ also represents the number of unknown parameters in model $M_n$, which is explicitly unrelated to the dimension term $p$. 
Therefore, our framework permits the number of nodes $p$ to increase rapidly, under the condition that both the number of candidate models $M_n$ and the maximum number of edges $k_{M_n}$ increase at a rate slower than that of the sample size $n$.


\begin{condition} \label{con:parameter2}
  There exists some positive constant $C$, such that $\max_{1 \le m \le M_n} \twonorm{\bA_*^{(m)}} \le C$ and $\max_{1 \le m \le M_n} k_{m}^{-1/2} \fnorm{\bA_*^{(m)}} \le C$.
\end{condition}

Condition~\ref{con:parameter2} limits the variability of the quasi-true values in candidate models. Specifically, it ensures that both the spectral norm and the Frobenius norm of these quasi-true coefficients are uniformly bounded from above.  A sufficient condition for this condition is that the elements of $\bA^{(m)}_*$ have scale of the order $O(1)$. This condition effectively controls the variance of the estimates, analogous to the constraint on the true coefficient in Condition~\ref{con:parameter}. A similar restriction condition on quasi-true values is also applied in the analysis of undirected Gaussian graphs, as discussed in \cite{liu2023Frequentist}.

\begin{condition} \label{con:riskformis}
  $\xi_n^{-1} \max \big\{(n k_{M_n} M_n)^{1/2}, \lambda_n k_{M_n} \big\} = o (1)$.
\end{condition}

Condition~\ref{con:riskformis} places restrictions on the rates at which the infimum Kullback-Leibler divergence $\xi_n$ and $\max \big\{(n k_{M_n} M_n)^{1/2}, \lambda_n k_{M_n} \big\}$ increase with the sample size $n$. A necessary condition for Condition~\ref{con:riskformis} is that $\xi_n \rightarrow \infty$, which implies that the candidate models should be substantially different from the correctly specified models. Specifically,  suppose that $i$\Th candidate model is correctly specified, and we set $\bw_i = 1$.  In this case, following the definition, $\KL_*(\bw)$ equals zero. Consequently, $\xi_n$ becomes zero when the set of candidate models includes a correctly specified model. Compared with Assumption 3 in \citet{liu2023Frequentist}, we allow the number of candidates $M_n$ diverges with the sample size $n$. 
Since $\lambda_n k_{M_n} $ is the upper bound of the additional penalty term, this condition illustrates that the effect of the penalty term $\lambda_n$ in the proposed criterion \eqref{eq:cirterion_penalized-likelihood} can be negligible with a proper value of $\lambda_n$, for example,  $\lambda_n=\log(n)$.


\begin{theorem}[Asymptotic optimality] \label{thm:asyopt}
When Conditions~\ref{con:parameter}, \ref{con:parameter2} and \ref{con:riskformis} are satisfied, it holds that
\begin{align*}
  \frac{\KL(\hbw)}{\inf _{\bw \in \mc{W}} \KL(\bw)} \stackrel{p}{\to} 1,
\end{align*}
where $\hbw$ is obtained by Equation \eqref{eq:weight}.
\end{theorem}
Theorem \ref{thm:asyopt} shows that our model averaging estimator $\hbA(\hbw)$ is optimal in the sense that its KL divergence is asymptotically identical to that of the theoretically best model averaging estimator.

\subsection{Asymptotic Properties of Model Weights}\label{limiting-property-of-the-model-weights}

Next, we present the asymptotic properties of the estimated weights tuned by \eqref{eq:weight}  when there is at least one correctly specified model in the set of DAGs. Before delving into the results, it is essential to introduce more detailed classifications of the candidate models to enable a comprehensive analysis of the variations among the models.  Without loss of generality, we assume that the first $M_0$ candidate models are underfitted, which miss some edges of the true model. These underfitted models are also categorized under misspecified models.  The remaining $M - M_0$ models are assumed to be correctly specified that contain all edges in the true model, where $0 \le M_0 < M_n$, as the candidate models are nested.  Let the $(M_0+1)$\Th candidate model be smallest correctly specified models, that has fewest edges of all correctly specified models. These remaining $M-M_0$ models can be  overfitted, containing additional edges not present in the true model.

Let $\hbw_{\mathrm{U}}$ and $\hbw_{\mathrm{O}}$ denote the subvectors  of $\hbw$ corresponding to the sets of underfitted models $\{1, \dots, M_0\}$, and overfitted models excluding the smallest correctly specified model $\{M_0+2, \dots, M_n\}$, respectively.
To derive the asymptotic property, we consider proving the following two parts: 
(1) the weights of the underfitted models converge to $0$;  
(2) the weights of overfitted models (except the smallest correctly specified models) converge to $0$.

We construct a new weight set as
\begin{align*}
  \mc{W}_{\mathrm{F}} = \Bigl\{ \bw \in [0,1]^{M_n}: \sum_{m=1}^{M_0} w_m = 1 \text{ and } \sum_{m=M_0+1}^{M_n} w_m = 0 \Bigr\} ,
\end{align*} 
where weights for  overfitted models are zeros. 
Let $\bD=(D_{ij}) \in\mathbb{R}^{M_0 \times M_0}$, where
\begin{align*}
   D_{ij} = \tr\Bigl\{ \bigl( {\widehat{\bA}}^{(i)} - {\widehat{\bA}}^{(M_0 + 1)} \bigr)\tran \bX\tran \bX\bigl( {\widehat{\bA}}^{(j)} - {\widehat{\bA}}^{(M_0 + 1)} \bigr) \Bigr\}, \text{ for } i,j\in\{1,\cdots,M_0\}.
\end{align*} 


\begin{condition} \label{con:diff_between_TF}
  There exists some positive constant $C$, such that $ s_{\min} (\bD) \ge C$ when $n$ is sufficiently large. 
\end{condition}
 
This condition controls the Frobenius inner product of differences between underfitted models and smallest correctly specified models. It ensures that the disparity among the underfitted models is sufficiently large, and there exists a non-negligible gap between the underfitted models and the smallest correctly specified model. This separation is crucial as it guarantees that the weights assigned to underfitted candidates will converge to zero.
Similar conditions are imposed in \cite{zouOptimalModelAveraging2022a} and \cite{liu2023Frequentist}.

\begin{theorem}[Weight consistency] \label{thm:weight_consistency}
  Under Conditions~\ref{con:parameter}, \ref{con:parameter2} and \ref{con:diff_between_TF},  when the $M_0+1$\Th model is the smallest correctly specified candidate model,
  the estimated weights with respect to underfitted models and overfitted models excluding the smallest correctly
specified model satisfy 
  \begin{align*}
    \| \hbw_{\mathrm{U}} \|_{2} = O_{p} \left( \lambda_{n}M_{0}^{1/2} {k_{M_0+1}}n^{- 1}  \right) \ \ \text{ and } \ \ 
    \| \hbw_{\mathrm{O}}\|_2 = O_{p} \left(  (M_n - M_0) k_{M_n} \lambda_{n}^{- 1}  \right),
  \end{align*}
  respectively.
\end{theorem}
Theorem \ref{thm:weight_consistency} demonstrates that when correctly specified models are included in the candidate set model, our proposed weight criterion will asymptotically assign zero weights to both underfitted and overfitted models, except for the smallest correctly specified model. Consequently, our method will asymptotically allocate all weights to the smallest correctly specified model. 
Note that the convergence rates of $\hbw_{\mathrm{U}}$ and  $\hbw_{\mathrm{O}}$ are related to the maximum number of edges in candidate models, other than explicitly related the dimension term $p$, which is used in \citet{liu2023Frequentist} for undirected graphs.  It is particularly suitable for scenarios where the candidate models are based on a small and predefined set of connected nodes. Specifically, the convergence rate of $\hbw_{\mathrm{U}}$ is proportional to the number of edges in the smallest correctly specified model, i.e., $k_{M_0+1}$, and is inversely proportional to the sample size $n$.  This means that as $n$ increases, the weights on underfitted models decrease sharply, with a smaller $k_{M_0+1}$ leading to a more rapid convergence of these weights to zeros. The convergence rate of $\hbw_{\mathrm{O}}$ is inversely proportional to $\lambda_n$. A constant $\lambda_n$ prevents $\hbw_{\mathrm{O}}$ from converging to zero. For example,  $\lambda_n = 2$ leads to the result that $\hbw_{\mathrm{O}}$ in the multivariate extension of Mallows model averaging does not converge to zero. This extends \citet{zhang2019inference}'s findings about Mallows model averaging where both $p$ and $M_n$ are fixed.

\subsection{Consistency of Model Averaging Estimator}

Next, we discuss the consistency of the model average estimator $\hbA(\hbw)$ when the candidate model set contains at least one correctly specified candidate model. Without loss of generality, we assume that the  $M_0+1$\Th model is the smallest correctly specified candidate model. Based on asymptotic properties of the estimated weights $\hbw$ in Theorem~\ref{thm:weight_consistency} and the consistency of each estimator $\hbA^{(m)}$ in Lemma~\ref{lemma:candidate_consistency}, we can deduce the consistency of  $\hbA(\hbw)$ as follows. 

\begin{theorem}[Parameter consistency] \label{thm:estimator_consistency}
  If Conditions~\ref{con:parameter}, \ref{con:parameter2} and \ref{con:diff_between_TF} hold,  when the $M_0+1$\Th model is the smallest correctly specified candidate model, we have
  \begin{align*}
    \ffnorm{\hbA(\hbw) - \bA_0} 
    = O_p \bigl( \lambda_{n}M_{0}^{1/2} {k_{M_0+1}^{3/2}} n^{- 1} \bigr) 
    + O_p \bigl( k_{M_0}^{1/2} n^{-1/2}\bigr) 
    + {O_p \bigl( (M_n - M_0)^{3/2} k_{M_n}^{3/2} \lambda_{n}^{- 1} n^{-1/2} \bigr)}
    .
  \end{align*}
\end{theorem}
Theorem~\ref{thm:estimator_consistency} shows the parameter consistency of the model averaging estimator $\hbA(\hbw)$. The three parts in right side of the equation in Theorem~\ref{thm:estimator_consistency} correspond to three categories of candidate models: underfitted models, the smallest correctly specified model, and overfitted models (except for the smallest correctly specified model). 
Importantly, Theorem~\ref{thm:estimator_consistency} illustrates that the convergence rate of the proposed estimator is aligned with the convergence speed of the smallest correctly specified model if $\lambda_n M_0^{1/2} k_{M_0+1} n^{-1/2} = o(1)$ and $(M_n - M_0)^{3/2} k_{M_n} \lambda_{n}^{- 1} = o(1)$.  The terms $\lambda_n M_0 k_{M_0+1} n^{-1/2}$ and $(M_n - M_0)^{3/2} k_{M_n} \lambda_{n}^{- 1}$ represent the prices we paid for the considering of uncertainty among underfitted and overfitted models, respectively. 



\section{Numerical Simulation}\label{numerical-study}

We generate directed acyclic graphs using the generation mechanisms introduced by \citet{kalisch2007estimating} and \citet{yuan2019constrained} with $p \in \{10,20 \}$. The sample size  $n$ is set from $50$ to $800$. For the construction of the true adjacency matrix $\bA_0$, we initially set $\bA_0 = \bzero$. Given a prespecified ordering of the $p$  nodes, we replace every entry on the lower off-diagonals of $\bA_0$ by a random sample of $0$ or $1$, which follows a Bernoulli distribution with the success probability $\rho \in \{0.2,0.5,0.8\}$. The number $1$ indicates an edge connection. Subsequently, we parameterize $\bA_0$ by replacing all entries with the value $1$ by  $0.5$. Given the generated $\bA_0$, random samples are generated according to Model \eqref{eq:model} with $\sigma = 1$.

To generate a sequence of nested candidate models, we use the forward and backward method inspired by the greedy equivalent search method \citep{chickering2002optimal} and the hill climbing method \citep{korb2010bayesian}. 
Specifically, we first obtain an initial graph selected by the constrained likelihood method with parameters tuned by the Hang-out method \citep{yuan2019constrained}. 
After the pre-selection, we apply a forward/backward method to incrementally add or remove edges, thereby generating a set of nested candidate models. 
The details have been discussed in Section \ref{model-averaging}. 
We set the number of candidate models by  $M_n=11$, with $5$ each derived from forward and backward methods.


We evaluate our proposed method (DAG-MA in figures below) against several benchmarks: the constrained likelihood method, the Bayesian model averaging approach utilizing a score-based Tabu search algorithm (BMA-Tabu) \citep{scutari2010bnlearn}, the Bayesian model averaging strategy based on a constraint-based Peter--Clark algorithm (BMA-PC) \citep{nagarajan2013bayesian}, and the Bayesian model averaging technique with the order Markov-Chain-Monte-Carlo learning algorithm (BMA-MCMC) \citep{suter2021BiDAG}. The first two Bayesian model averaging methods are implemented in the R package \emph{bnlearn}, and the third one in the R package \emph{BiDAG}.

To exam the performance of different methods, we measure the accuracy of parameter estimation by the Kullback-Leibler (KL) loss to measure the discrepancy between the true inverse covariance matrix $\bOmg_{0} = (\bI-\bA_0) (\bI-\bA_0\t) / \sigma^2$ and its estimate $\widehat{\bOmg}$ with using the estimates of $\bA_0$ and $\sigma^2$:
\begin{align*}
  \operatorname{KL}\left(\widehat{\bOmg}, \bOmg_{0}\right)=\operatorname{tr}\left(\bOmg_{0}^{-1} \widehat{\bOmg}\right)-\log \left|\bOmg_{0}^{-1} \widehat{\bOmg}\right|-p.
\end{align*}
 The prediction accuracy is measured by prediction error (PE),
\begin{align*}
  \operatorname{PE}(\hbA) = \frac{1}{np} \fnorm{\bX\bA_0 - \bX \hbA}.
\end{align*}
We also exam the estimation errors (EE) of $\bA_0$ and $\bOmg_0$ by
\begin{align*}
  \operatorname{EE}(\hbA) = \fnorm{\bA_0 - \hbA}  \text{ and } \operatorname{EE}(\hbOmg) = \fnorm{\bOmg_0 - \hbOmg},
\end{align*}
respectively.

\begin{figure} 
  \centering
  \caption{Average KL divergences of estimates of competing methods as sample size increases under different settings for $(p, \rho)$.}
  \label{fig:KL}

 \begin{subfigure}{0.32\linewidth}
    \centering
    \includegraphics[width=\linewidth]{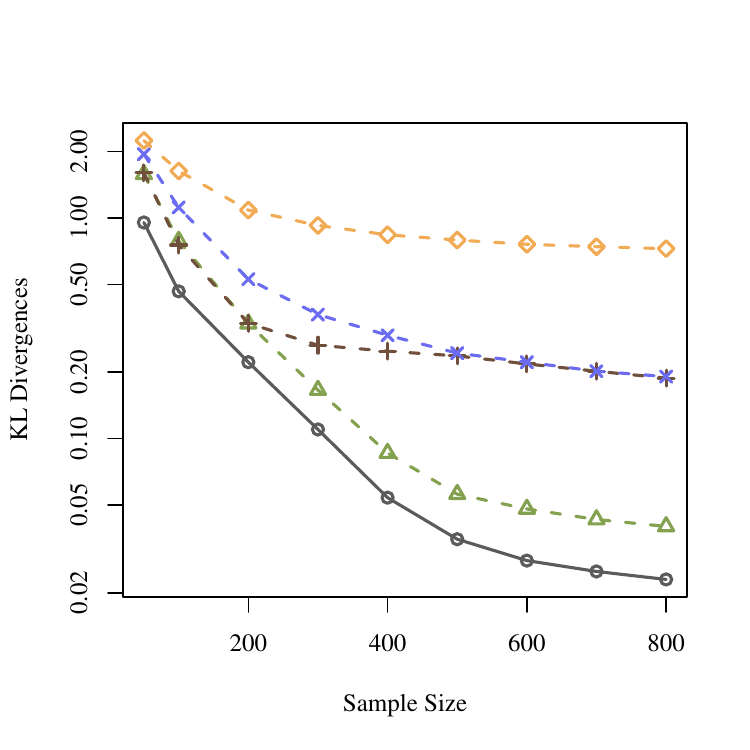}
    \caption{$p=10$, $\rho=0.2$}
    \label{subfig:klp10rho2}
  \end{subfigure}\hfill
  \begin{subfigure}{0.32\linewidth}
    \centering
    \includegraphics[width=\linewidth]{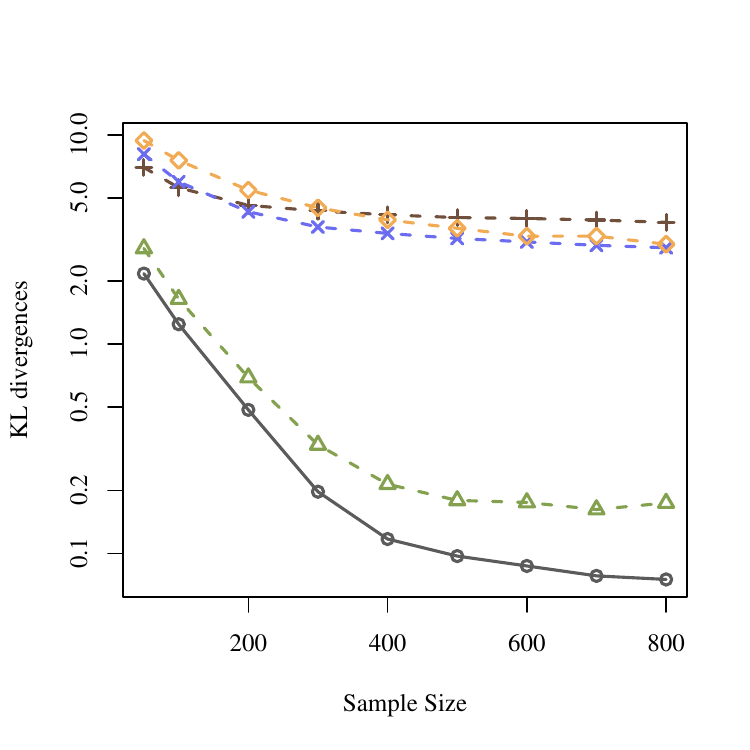}
    \caption{$p=10$, $\rho=0.5$}
    \label{subfig:klp10rho5}
  \end{subfigure}\hfill
  \begin{subfigure}{0.32\linewidth}
    \centering
    \includegraphics[width=\linewidth]{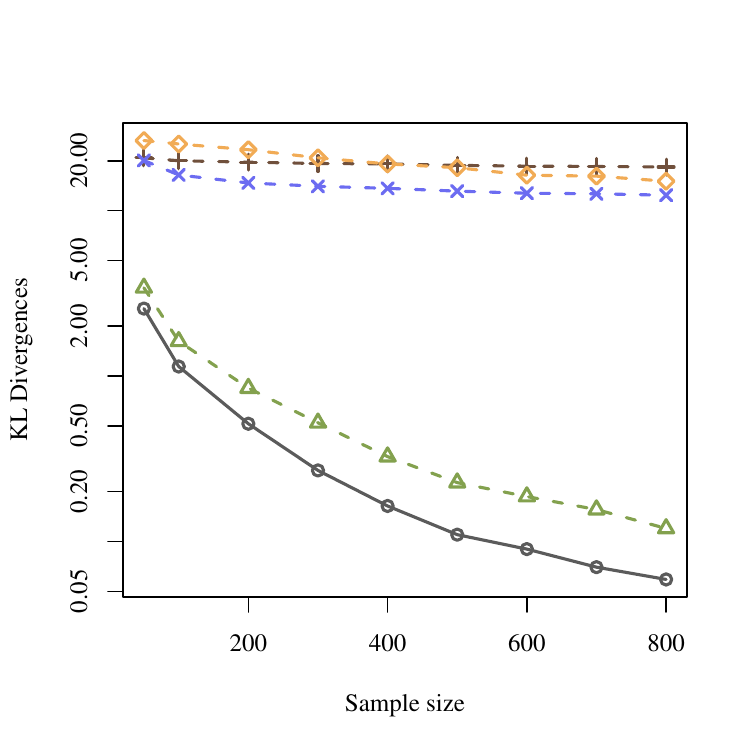}
    \caption{$p=10$, $\rho=0.8$}
    \label{subfig:klp10rho8}
  \end{subfigure}

  \medskip

  \begin{subfigure}{0.32\linewidth}
    \centering
    \includegraphics[width=\linewidth]{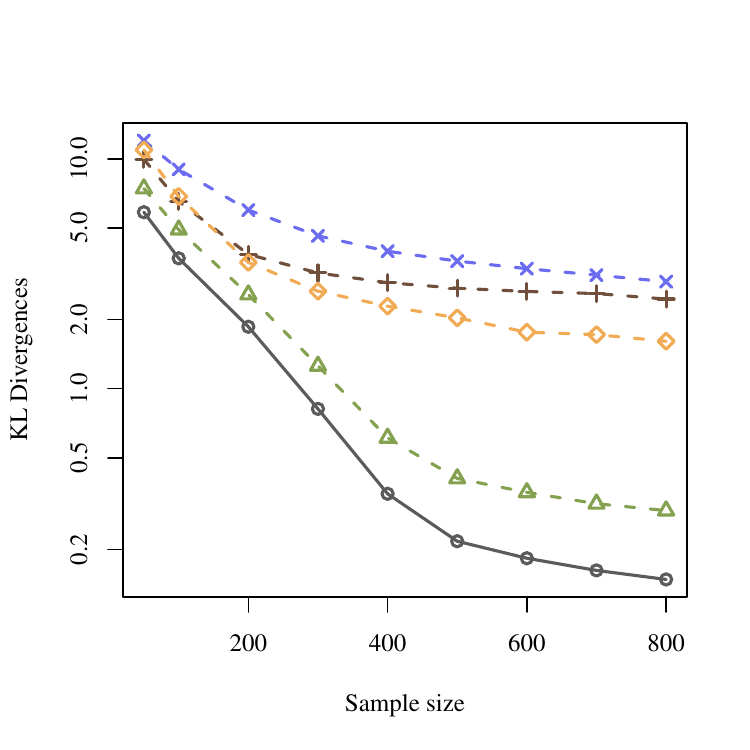}
    \caption{$p=20$, $\rho=0.2$}
    \label{subfig:klp20rho2}
  \end{subfigure}\hfill
  \begin{subfigure}{0.32\linewidth}
    \centering
    \includegraphics[width=\linewidth]{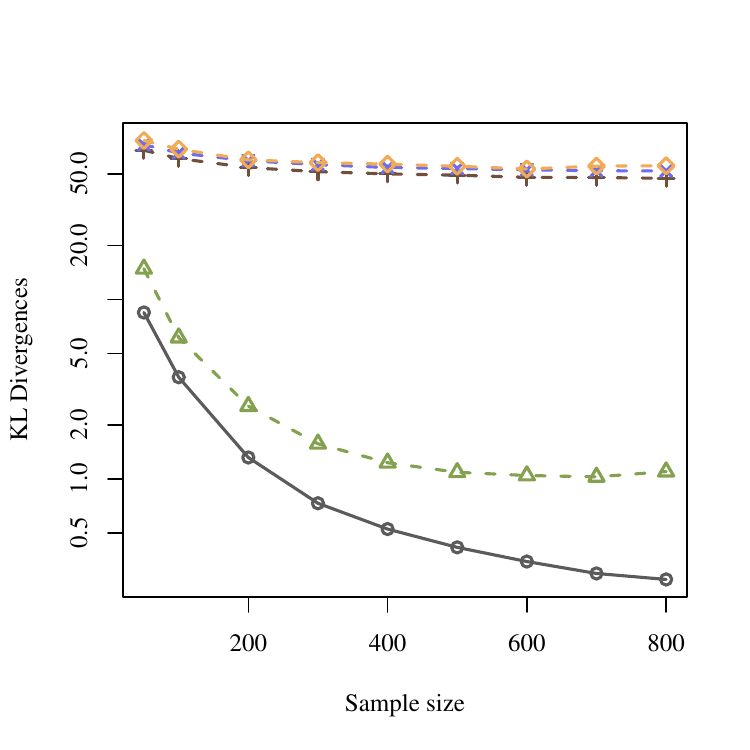}
    \caption{$p=20$, $\rho=0.5$}
    \label{subfig:klp20rho5}
  \end{subfigure}\hfill
  \begin{subfigure}{0.32\linewidth}
    \centering
    \includegraphics[width=\linewidth]{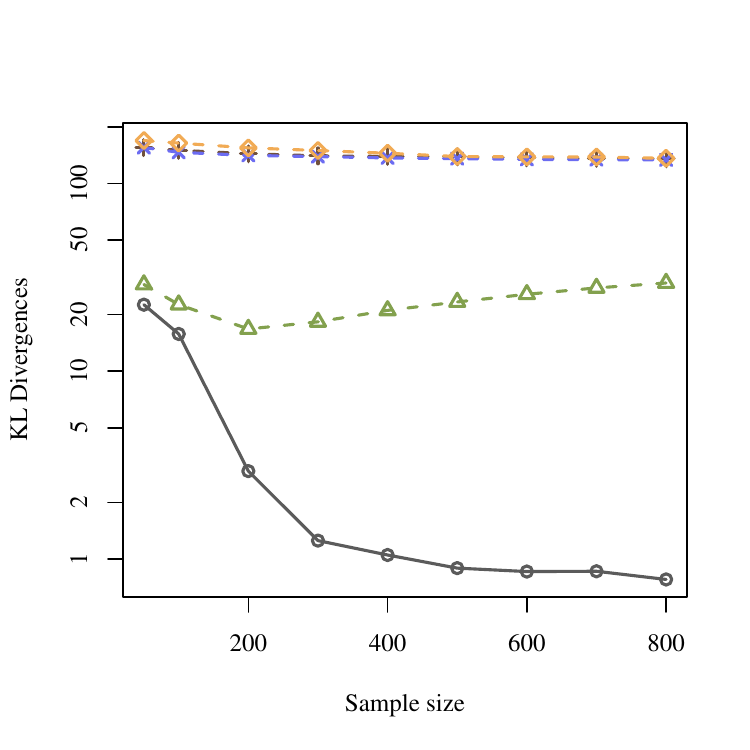}
    \caption{$p=20$, $\rho=0.8$}
    \label{subfig:klp20rho8}
  \end{subfigure}
 
\vspace{0.4em}
\begin{center}
\begin{tikzpicture}[x=1cm,y=1cm]
  \def\dx{4.0}   
  \def\dy{0.9}   
  \def\L{0.9}    
  \def\m{0.45}   
  \def\labx{1.3} 

  \draw[black,thick] (0,0) -- (\L,0);
  \draw[black,thick] (\m,0) circle (2pt);
  \node[anchor=west] at (\labx,0) {DAG-MA};

  \draw[mygreen,dashed,thick] (\dx,0) -- ++(\L,0);
  \node at (\dx+\m,0) {\textcolor{mygreen}{\footnotesize$\triangle$}};
  \node[anchor=west] at (\dx+\labx,0) {Constrained Likelihood Method};

  \draw[mybrown,dashed,thick] (0,-\dy) -- (\L,-\dy);
  \node at (\m,-\dy) {\textcolor{mybrown}{\footnotesize +}};
  \node[anchor=west] at (\labx,-\dy) {BMA-Tabu};

  \draw[mypurple,dashed,thick] (\dx,-\dy) -- ++(\L,0);
  \node at (\dx+\m,-\dy) {\textcolor{mypurple}{\footnotesize $\times$}};
  \node[anchor=west] at (\dx+\labx,-\dy) {BMA-PC};

  \draw[myorange,dashed,thick] (2*\dx,-\dy) -- ++(\L,0);
  \node at (2*\dx+\m,-\dy) {\textcolor{myorange}{$\diamond$}};
  \node[anchor=west] at (2*\dx+\labx,-\dy) {BMA-MCMC};
\end{tikzpicture}
\end{center}

  \begin{minipage}{\linewidth}
    \footnotesize \emph{Notes.} The figure plots average KL divergences under different numbers of nodes $p$ and success probabilities $\rho$ of the Bernoulli distributions generating the lower off-diagonal elements of the true adjacency matrix. The proposed DAG-MA method is compared with four benchmarks: the constrained likelihood method, BMA with Tabu search (BMA-Tabu), BMA with the Peter--Clark algorithm (BMA-PC), and BMA with order MCMC (BMA-MCMC). 
  \end{minipage}
\end{figure}
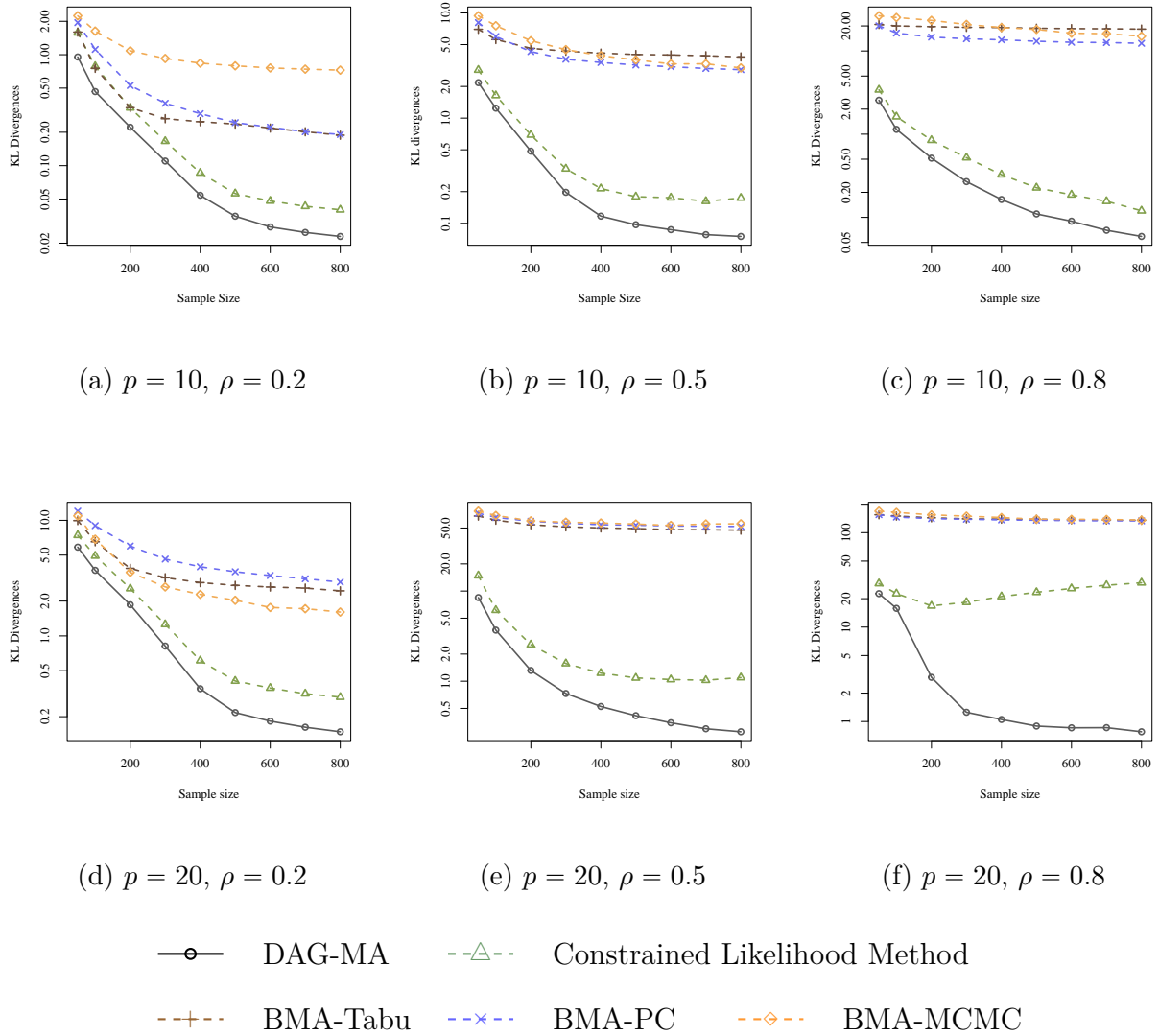

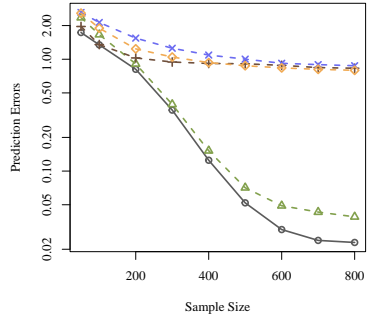
\begin{figure} 
 \caption{Average prediction errors of competing methods as sample size increases under different settings for $(p, \rho)$. }
  \label{fig:PE}
  \centering
\begin{subfigure}{0.32\linewidth}
  \centering
  \includegraphics[width=\linewidth]{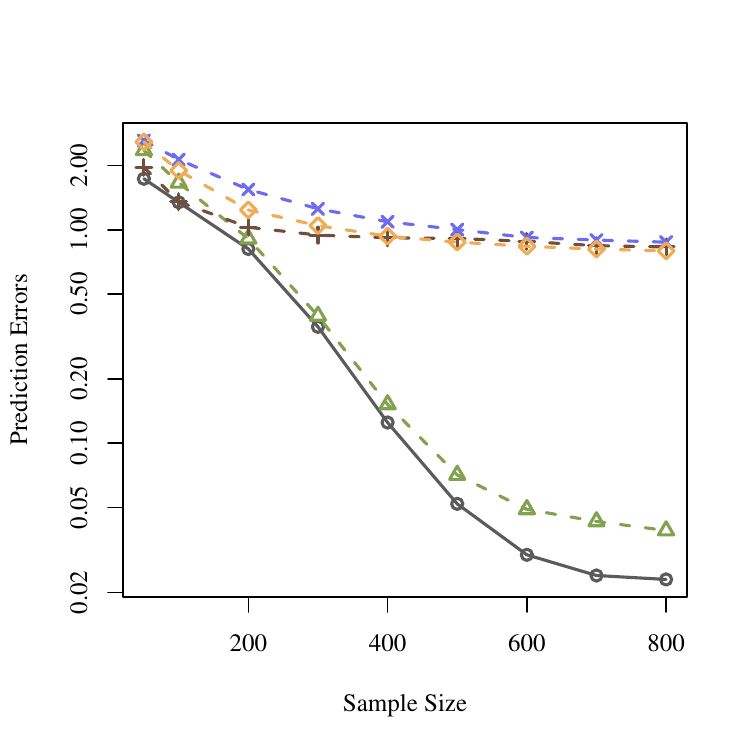}
  \caption{$p=10$, $\rho=0.2$}
  \label{subfig:pep10rho2}
\end{subfigure}\hfill
\begin{subfigure}{0.32\linewidth}
  \centering
  \includegraphics[width=\linewidth]{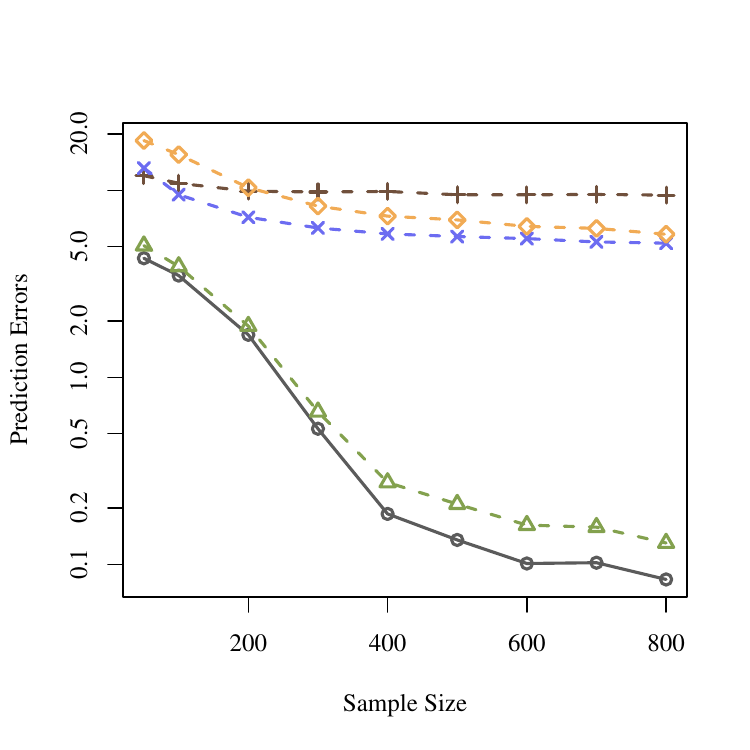}
  \caption{$p=10$, $\rho=0.5$}
  \label{subfig:pep10rho5}
\end{subfigure}\hfill
\begin{subfigure}{0.32\linewidth}
  \centering
  \includegraphics[width=\linewidth]{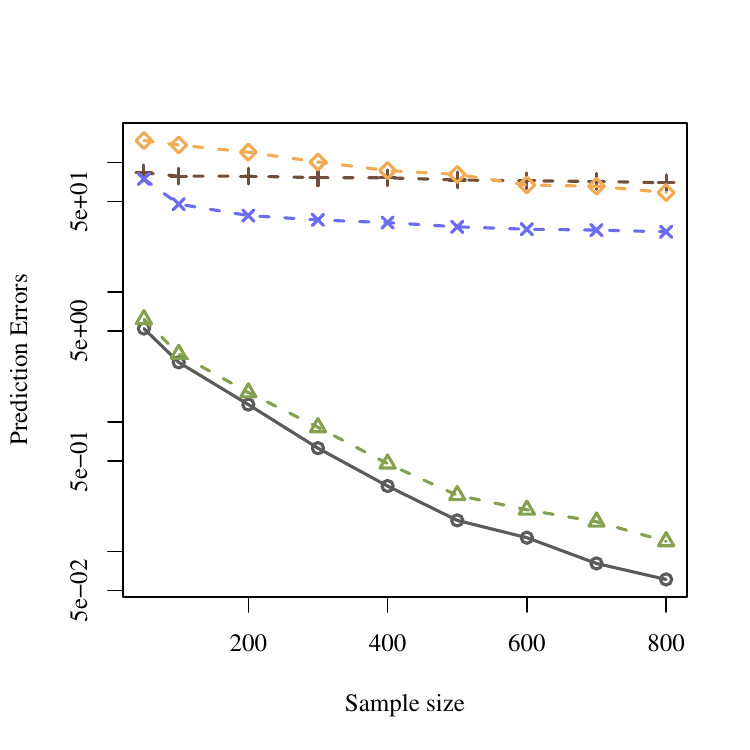}
  \caption{$p=10$, $\rho=0.8$}
  \label{subfig:pep10rho8}
\end{subfigure}

\medskip

\begin{subfigure}{0.32\linewidth}
  \centering
  \includegraphics[width=\linewidth]{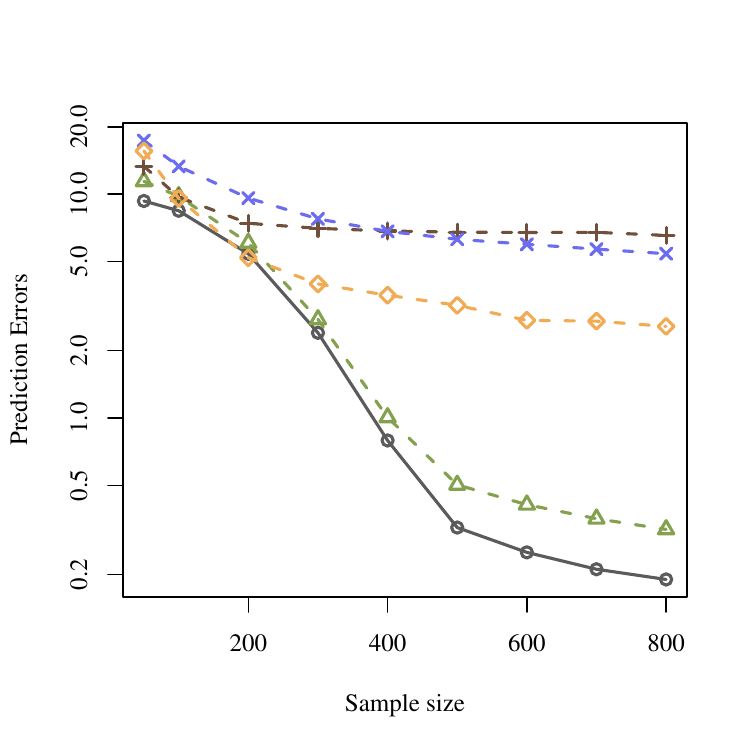}
  \caption{$p=20$, $\rho=0.2$}
  \label{subfig:pep20rho2}
\end{subfigure}\hfill
\begin{subfigure}{0.32\linewidth}
  \centering
  \includegraphics[width=\linewidth]{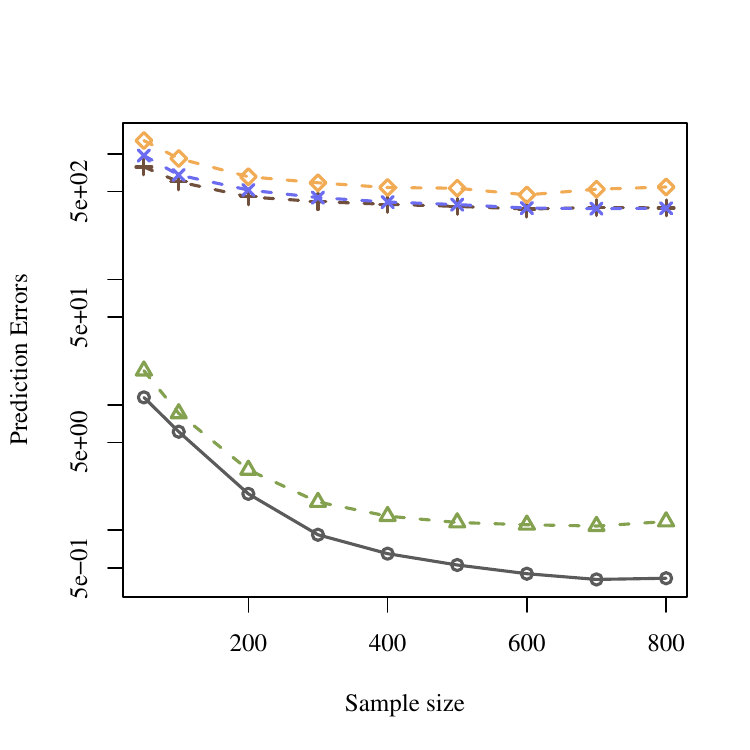}
  \caption{$p=20$, $\rho=0.5$}
  \label{subfig:pep20rho5}
\end{subfigure}\hfill
\begin{subfigure}{0.32\linewidth}
  \centering
  \includegraphics[width=\linewidth]{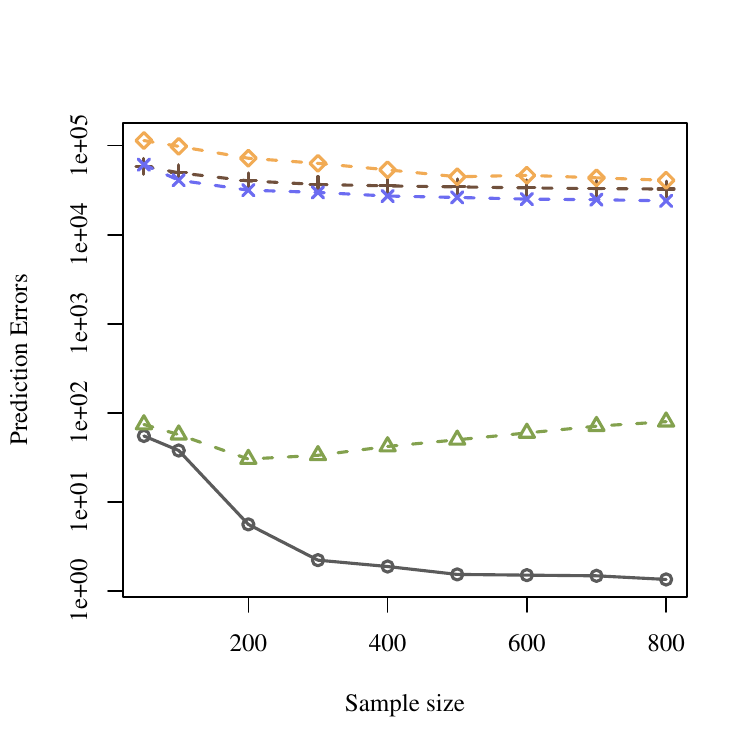}
  \caption{$p=20$, $\rho=0.8$}
  \label{subfig:pep20rho8}
\end{subfigure}

\vspace{0.4em}
\begin{center}
\begin{tikzpicture}[x=1cm,y=1cm]
  \def\dx{4.0}   
  \def\dy{0.9}   
  \def\L{0.9}    
  \def\m{0.45}   
  \def\labx{1.3} 

  \draw[black,thick] (0,0) -- (\L,0);
  \draw[black,thick] (\m,0) circle (2pt);
  \node[anchor=west] at (\labx,0) {DAG-MA};

  \draw[mygreen,dashed,thick] (\dx,0) -- ++(\L,0);
  \node at (\dx+\m,0) {\textcolor{mygreen}{\footnotesize$\triangle$}};
  \node[anchor=west] at (\dx+\labx,0) {Constrained Likelihood Method};

  \draw[mybrown,dashed,thick] (0,-\dy) -- (\L,-\dy);
  \node at (\m,-\dy) {\textcolor{mybrown}{\footnotesize +}};
  \node[anchor=west] at (\labx,-\dy) {BMA-Tabu};

  \draw[mypurple,dashed,thick] (\dx,-\dy) -- ++(\L,0);
  \node at (\dx+\m,-\dy) {\textcolor{mypurple}{\footnotesize $\times$}};
  \node[anchor=west] at (\dx+\labx,-\dy) {BMA-PC};

  \draw[myorange,dashed,thick] (2*\dx,-\dy) -- ++(\L,0);
  \node at (2*\dx+\m,-\dy) {\textcolor{myorange}{$\diamond$}};
  \node[anchor=west] at (2*\dx+\labx,-\dy) {BMA-MCMC};
\end{tikzpicture}
\end{center}

  \begin{minipage}{\linewidth}
    \footnotesize \emph{Notes.} The figure plots average prediction errors under different numbers of nodes $p$ and success probabilities $\rho$ of the Bernoulli distributions generating the lower off-diagonal elements of the true adjacency matrix. The proposed DAG-MA method is compared with four benchmarks: the constrained likelihood method, BMA with Tabu search (BMA-Tabu), BMA with the Peter--Clark algorithm (BMA-PC), and BMA with order MCMC (BMA-MCMC). 
  \end{minipage}
\end{figure}

\begin{figure}
 \caption{Average estimation errors for $\bA_0$ of competing methods as sample size increases under different settings for $(p, \rho)$. }
  \label{fig:eeA}
  \centering 
  \begin{subfigure}{0.32\linewidth}
  \centering
  \includegraphics[width=\linewidth]{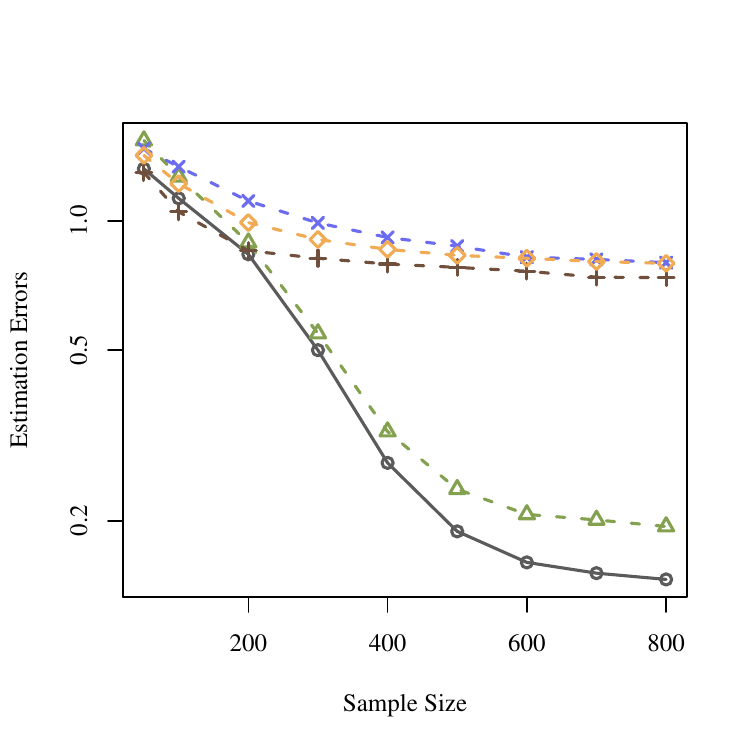}
  \caption{$p=10$, $\rho=0.2$}
  \label{subfig:eeAp10rho2}
\end{subfigure}\hfill
\begin{subfigure}{0.32\linewidth}
  \centering
  \includegraphics[width=\linewidth]{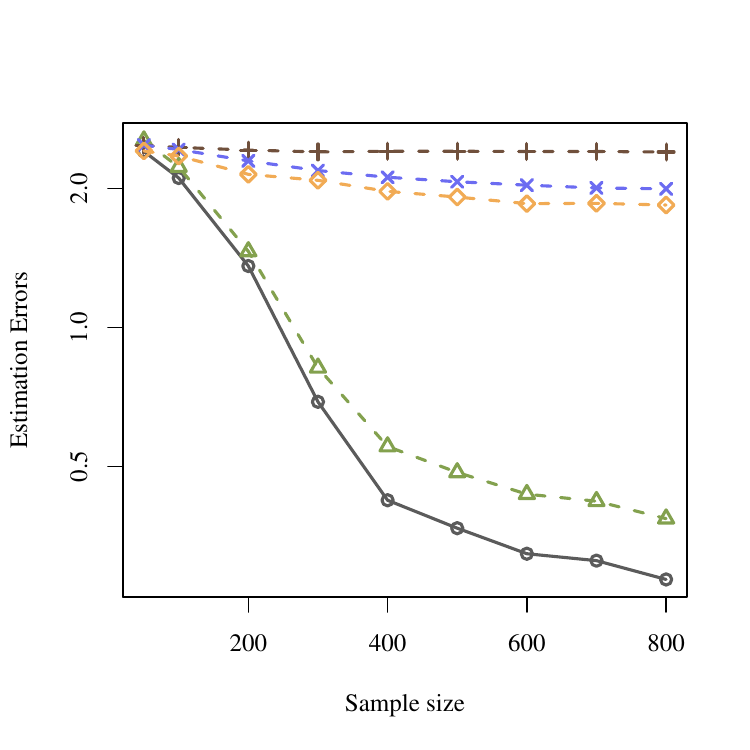}
  \caption{$p=10$, $\rho=0.5$}
  \label{subfig:eeAp10rho5}
\end{subfigure}\hfill
\begin{subfigure}{0.32\linewidth}
  \centering
  \includegraphics[width=\linewidth]{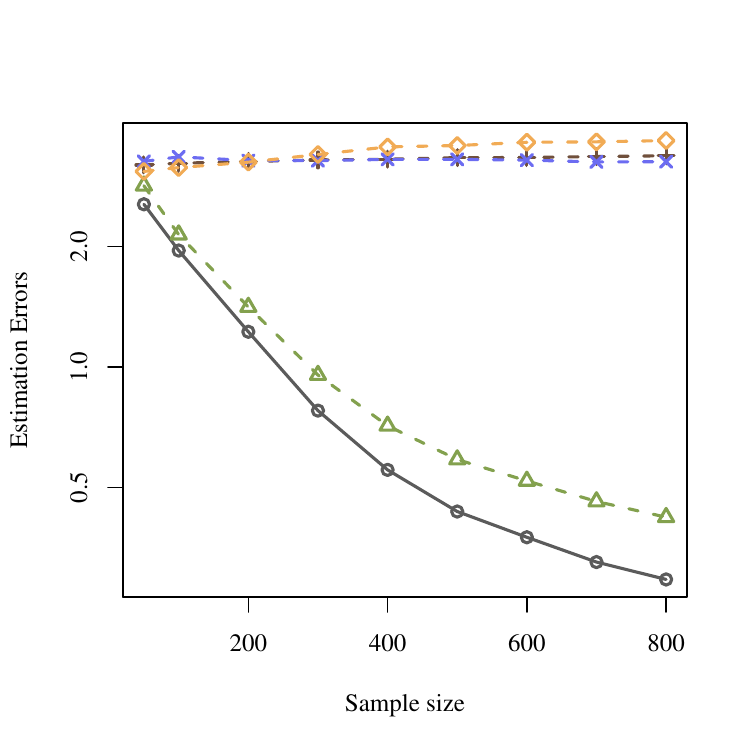}
  \caption{$p=10$, $\rho=0.8$}
  \label{subfig:eeAp10rho8}
\end{subfigure}

\medskip

\begin{subfigure}{0.32\linewidth}
  \centering
  \includegraphics[width=\linewidth]{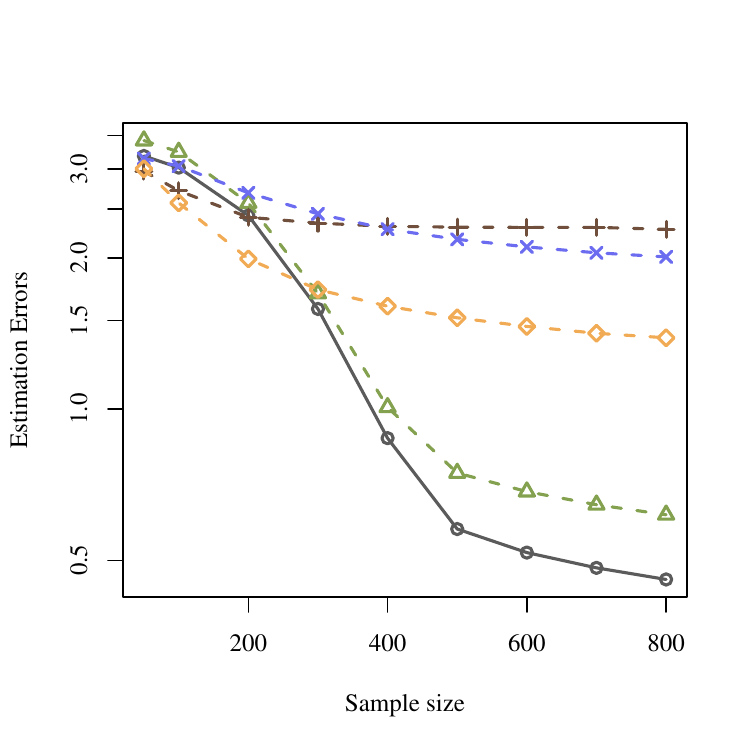}
  \caption{$p=20$, $\rho=0.2$}
  \label{subfig:eeAp20rho2}
\end{subfigure}\hfill
\begin{subfigure}{0.32\linewidth}
  \centering
  \includegraphics[width=\linewidth]{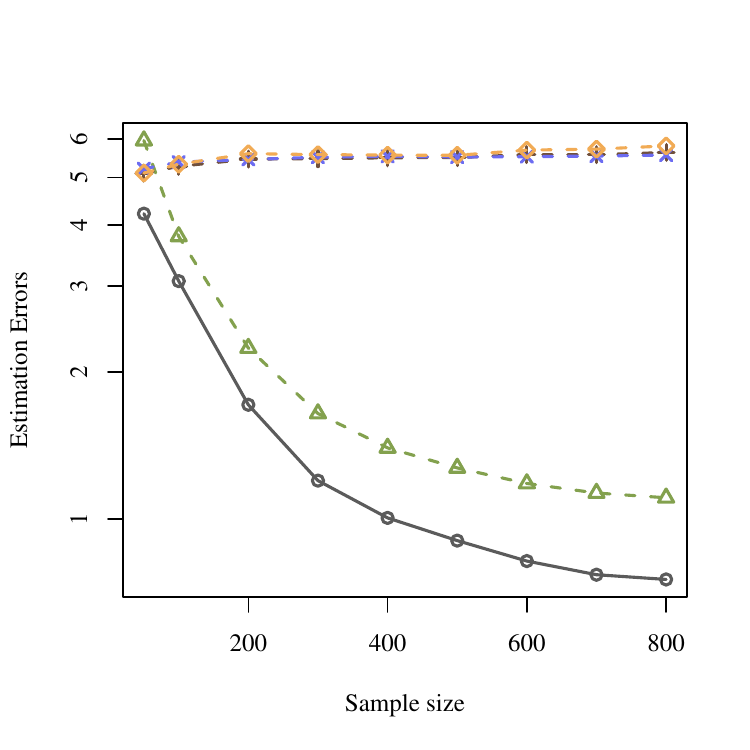}
  \caption{$p=20$, $\rho=0.5$}
  \label{subfig:eeAp20rho5}
\end{subfigure}\hfill
\begin{subfigure}{0.32\linewidth}
  \centering
  \includegraphics[width=\linewidth]{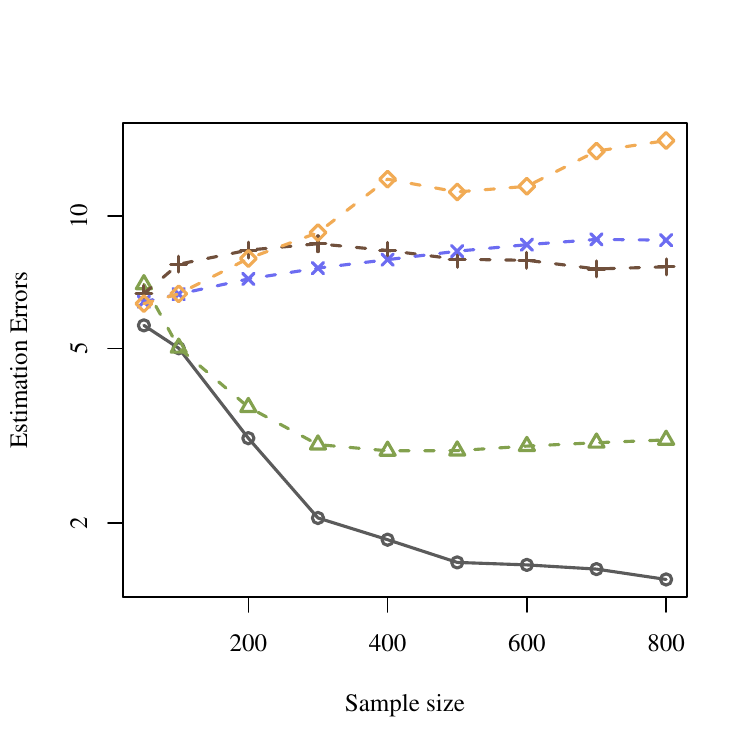}
  \caption{$p=20$, $\rho=0.8$}
  \label{subfig:eeAp20rho8}
\end{subfigure}

\vspace{0.4em}
\begin{center}
\begin{tikzpicture}[x=1cm,y=1cm]
  \def\dx{4.0}   
  \def\dy{0.9}   
  \def\L{0.9}    
  \def\m{0.45}   
  \def\labx{1.3} 

  \draw[black,thick] (0,0) -- (\L,0);
  \draw[black,thick] (\m,0) circle (2pt);
  \node[anchor=west] at (\labx,0) {DAG-MA};

  \draw[mygreen,dashed,thick] (\dx,0) -- ++(\L,0);
  \node at (\dx+\m,0) {\textcolor{mygreen}{\footnotesize$\triangle$}};
  \node[anchor=west] at (\dx+\labx,0) {Constrained Likelihood Method};

  \draw[mybrown,dashed,thick] (0,-\dy) -- (\L,-\dy);
  \node at (\m,-\dy) {\textcolor{mybrown}{\footnotesize +}};
  \node[anchor=west] at (\labx,-\dy) {BMA-Tabu};

  \draw[mypurple,dashed,thick] (\dx,-\dy) -- ++(\L,0);
  \node at (\dx+\m,-\dy) {\textcolor{mypurple}{\footnotesize $\times$}};
  \node[anchor=west] at (\dx+\labx,-\dy) {BMA-PC};

  \draw[myorange,dashed,thick] (2*\dx,-\dy) -- ++(\L,0);
  \node at (2*\dx+\m,-\dy) {\textcolor{myorange}{$\diamond$}};
  \node[anchor=west] at (2*\dx+\labx,-\dy) {BMA-MCMC};
\end{tikzpicture}
\end{center}

  \begin{minipage}{\linewidth}
    \footnotesize \emph{Notes.} The figure plots average estimation errors for estimated  adjacency matrix under different numbers of nodes $p$ and success probabilities $\rho$ of the Bernoulli distributions generating the lower off-diagonal elements of the true adjacency matrix. The proposed DAG-MA method is compared with four benchmarks: the constrained likelihood method, BMA with Tabu search (BMA-Tabu), BMA with the Peter--Clark algorithm (BMA-PC), and BMA with order MCMC (BMA-MCMC). 
  \end{minipage}
\end{figure}

\begin{figure} 
 \caption{Average estimation errors for $\bOmg_0$ of competing methods as sample size increases under different settings for $(p, \rho)$.}
  \label{fig:eeOmg}
  \centering
\begin{subfigure}{0.32\linewidth}
  \centering
  \includegraphics[width=\linewidth]{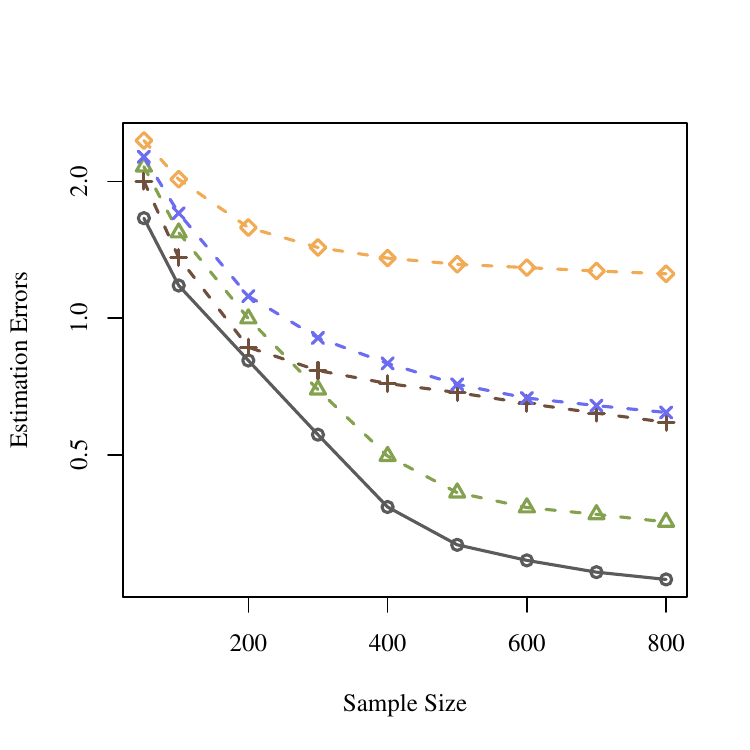}
  \caption{$p=10$, $\rho=0.2$}
  \label{subfig:eeOmgp10rho2}
\end{subfigure}\hfill
\begin{subfigure}{0.32\linewidth}
  \centering
  \includegraphics[width=\linewidth]{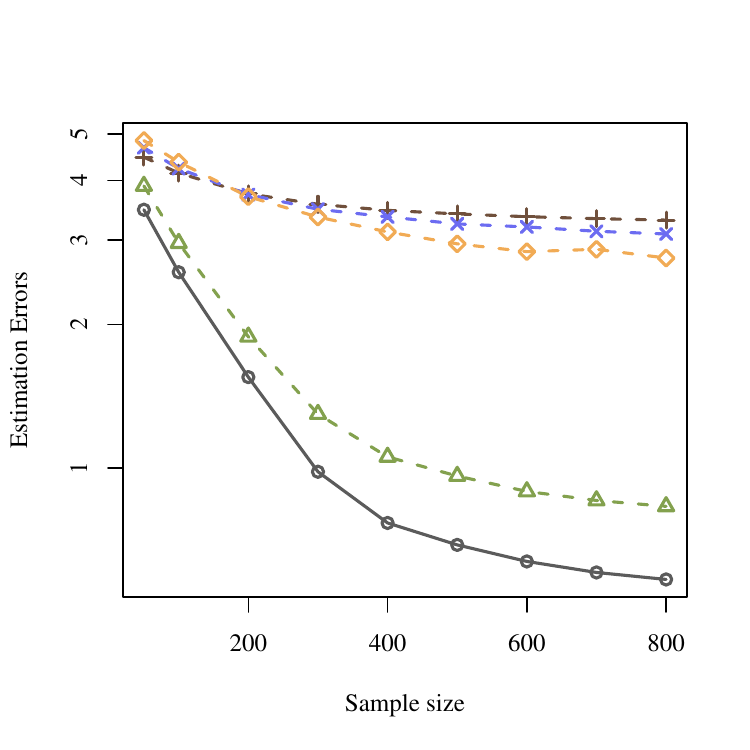}
  \caption{$p=10$, $\rho=0.5$}
  \label{subfig:eeOmgp10rho5}
\end{subfigure}\hfill
\begin{subfigure}{0.32\linewidth}
  \centering
  \includegraphics[width=\linewidth]{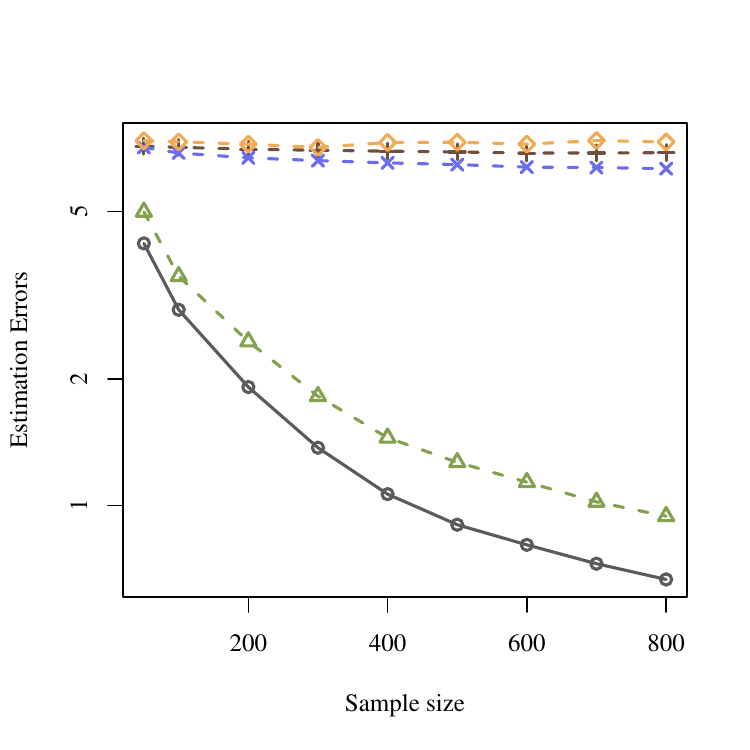}
  \caption{$p=10$, $\rho=0.8$}
  \label{subfig:eeOmgp10rho8}
\end{subfigure}

\medskip

\begin{subfigure}{0.32\linewidth}
  \centering
  \includegraphics[width=\linewidth]{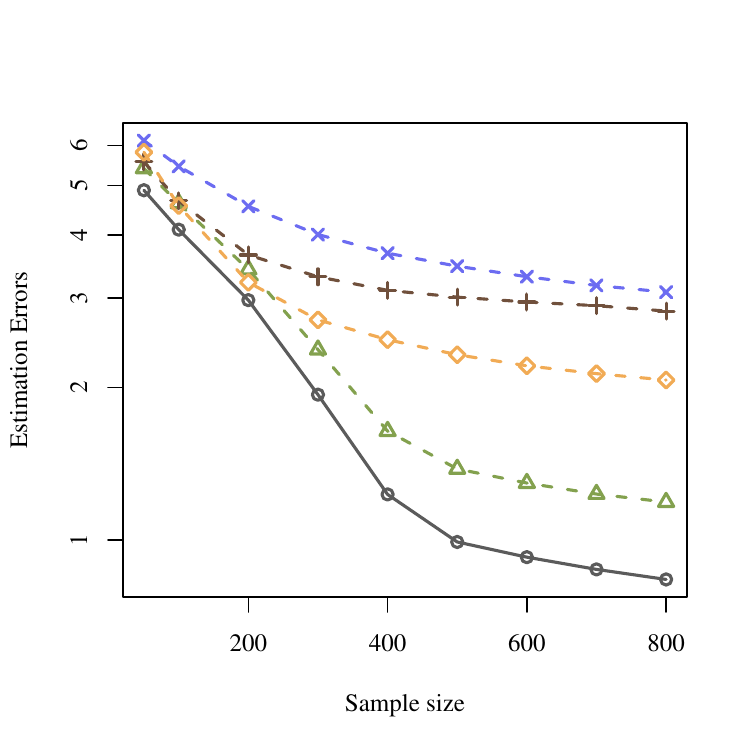}
  \caption{$p=20$, $\rho=0.2$}
  \label{subfig:eeOmgp20rho2}
\end{subfigure}\hfill
\begin{subfigure}{0.32\linewidth}
  \centering
  \includegraphics[width=\linewidth]{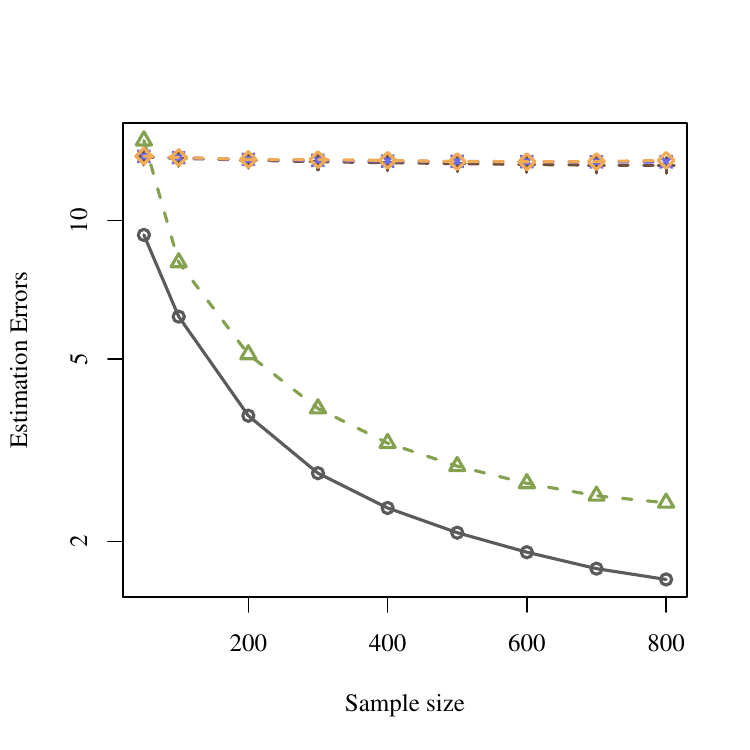}
  \caption{$p=20$, $\rho=0.5$}
  \label{subfig:eeOmgp20rho5}
\end{subfigure}\hfill
\begin{subfigure}{0.32\linewidth}
  \centering
  \includegraphics[width=\linewidth]{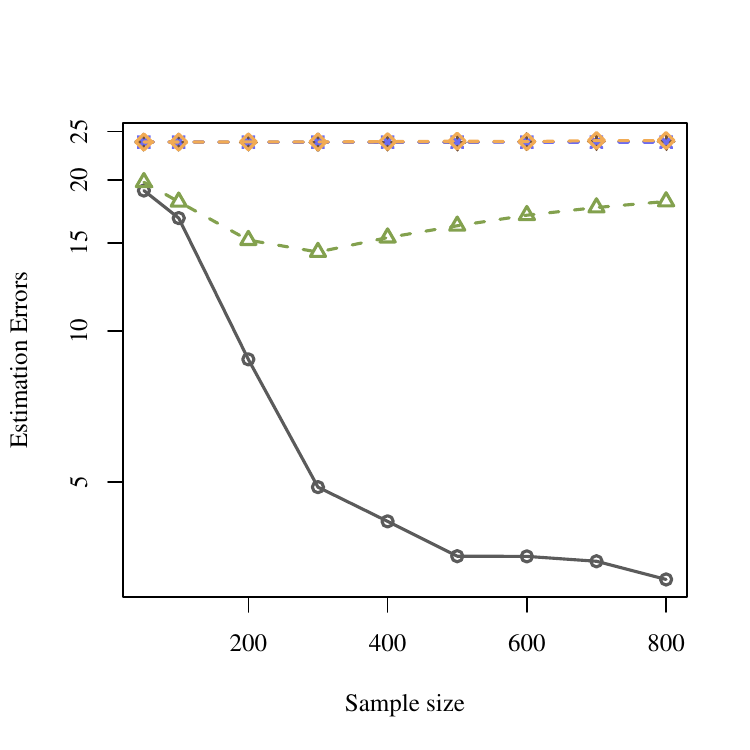}
  \caption{$p=20$, $\rho=0.8$}
  \label{subfig:eeOmgp20rho8}
\end{subfigure}

\vspace{0.4em}
\begin{center}
\begin{tikzpicture}[x=1cm,y=1cm]
  \def\dx{4.0}   
  \def\dy{0.9}   
  \def\L{0.9}    
  \def\m{0.45}   
  \def\labx{1.3} 

  \draw[black,thick] (0,0) -- (\L,0);
  \draw[black,thick] (\m,0) circle (2pt);
  \node[anchor=west] at (\labx,0) {DAG-MA};

  \draw[mygreen,dashed,thick] (\dx,0) -- ++(\L,0);
  \node at (\dx+\m,0) {\textcolor{mygreen}{\footnotesize$\triangle$}};
  \node[anchor=west] at (\dx+\labx,0) {Constrained Likelihood Method};

  \draw[mybrown,dashed,thick] (0,-\dy) -- (\L,-\dy);
  \node at (\m,-\dy) {\textcolor{mybrown}{\footnotesize +}};
  \node[anchor=west] at (\labx,-\dy) {BMA-Tabu};

  \draw[mypurple,dashed,thick] (\dx,-\dy) -- ++(\L,0);
  \node at (\dx+\m,-\dy) {\textcolor{mypurple}{\footnotesize $\times$}};
  \node[anchor=west] at (\dx+\labx,-\dy) {BMA-PC};

  \draw[myorange,dashed,thick] (2*\dx,-\dy) -- ++(\L,0);
  \node at (2*\dx+\m,-\dy) {\textcolor{myorange}{$\diamond$}};
  \node[anchor=west] at (2*\dx+\labx,-\dy) {BMA-MCMC};
\end{tikzpicture}
\end{center}

\begin{minipage}{\linewidth}
    \footnotesize \emph{Notes.} The figure plots average estimation errors for estimated  inverse covariance matrix under different numbers of nodes $p$ and success probabilities $\rho$ of the Bernoulli distributions generating the lower off-diagonal elements of the true adjacency matrix. The proposed DAG-MA method is compared with four benchmarks: the constrained likelihood method, BMA with Tabu search (BMA-Tabu), BMA with the Peter--Clark algorithm (BMA-PC), and BMA with order MCMC (BMA-MCMC). 
\end{minipage}

\end{figure}

For each setting of $n$, $p$, and $\rho$, we generate $500$ replications and calculate the average values for each evaluation criterion. We present the results of the above four criteria of competing methods under different pairs of $(p,\rho)$ in Fig.\ref{fig:KL}--Fig.\ref{fig:eeOmg}, respectively, where the vertical axes of these figures  have logarithmic scale with base-$10$ for better presentation.   

The results depicted in Fig.\ref{fig:KL} indicate that our method outperforms competing models, achieving the lowest average Kullback-Leibler divergences. This advantage is more pronounced for larger sample sizes. From Fig.\ref{fig:PE}, it is evident that our method consistently attains the lowest average prediction errors in the majority of cases. Fig.\ref{fig:eeA} reveals that the average estimation errors of $\bA_0$ of our method are the lowest across most scenarios and that these errors diminish as the sample size grows for all pairs of $(p, \rho)$, aligning with the outcomes stated in Theorem \ref{thm:estimator_consistency}. Furthermore, Fig.\ref{fig:eeOmg} demonstrates that our method has the lowest average estimation errors  of $\hbOmg_0$ in all examined cases.

\section{Analysis of Locational Banking Statistics (LBS)} \label{real-data}

Numerous studies have explored the relationship among countries (or regions) in the cross-border banking market \citep{Bremus2015JBF,TONZER2015JFS,giudici2016Graphical}. However, there is still skepticism regarding the contemporaneous relationship between these banking countries (or regions) and a limited understanding of the integration of the international banking market. We analyze the relationships among 26 countries (or regions) using datasets of aggregate cross-border exposure recorded by LBS \footnote{The data is available in https://data.bis.org/topics/LBS}. The list of these 26 countries (or regions) is detailed in Table~\ref{tab:real_data} and encompasses the countries (or regions) discussed in \citet{giudici2016Graphical}.  Each country (or region) is represented by the value of its liabilities toward all other countries, measured on a quarterly basis from the third quarter of 1997 (Q3 1997) to the first quarter of 2022 (Q1 2022).  We standardize the data by rescaling the observed values at each timestamp to achieve zero mean and unit variance.

\clearpage
\begin{table} 
  \caption{Countries and regions included in the analysis} 
  \label{tab:real_data}
  \begin{center}
       
  \begin{tabular}{p{2cm}p{5cm}p{2cm}p{5cm}}
    \toprule
    Code & Country / Region & Code & Country / Region \\
    \midrule
    AT & Austria        & IT & Italy                \\
    BS & Bahamas        & JP & Japan                \\
    BH & Bahrain        & LU & Luxembourg            \\
    BE & Belgium        & NL & Netherlands          \\
    CA & Canada         & NO & Norway               \\
    KY & Cayman Islands & SG & Singapore            \\
    DK & Denmark        & ES & Spain                \\
    FI & Finland        & SE & Sweden               \\
    FR & France         & CH & Switzerland          \\
    DE & Germany        & GB & United Kingdom       \\
    HK & Hong Kong SAR  & US & United States        \\
    IE & Ireland        & CN & China                \\
    IN & India          & KR & South Korea          \\
    \bottomrule
  \end{tabular}
  \end{center}
\vspace{0.3em}
\footnotesize \emph{Notes.} The table lists the 26 countries and regions included in the empirical analysis, with their corresponding codes. The sample is constructed from LBS, covering quarterly cross-border banking liabilities from Q3 1997 to Q1 2022.
\end{table}
\clearpage
 
\begin{figure} 
  \caption{
    Estimated DAG learned by DAG-MA for LBS.
    } 
    \label{fig:real_data}
    \begin{center}
    \includegraphics[width=\linewidth]{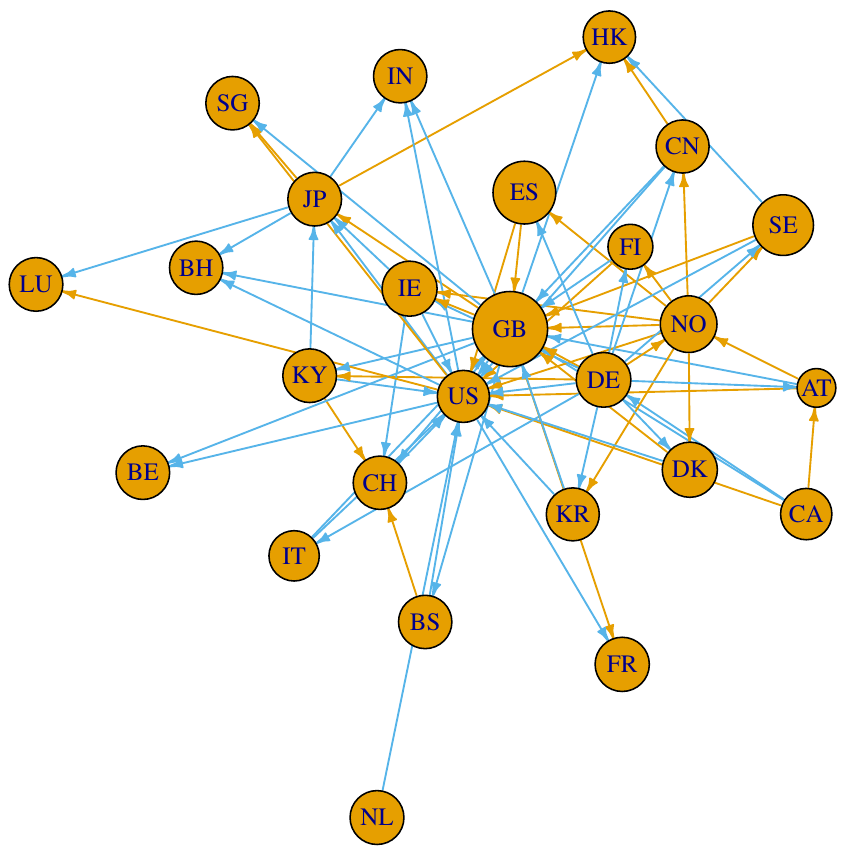}
    \end{center}
    \par
 \vspace{0.5em}
  \footnotesize \emph{Notes.} The figure shows the DAG learned from LBS data by the proposed method. The arrows represent causal effects, and the color of the arrows indicates the positive or negative effect (blue is negative, orange is positive).
\end{figure}

We employ our proposed method on the LBS data and obtain the resulting estimated graph, which can be viewed in Fig.\ref{fig:real_data}. From Fig.\ref{fig:real_data}, a number of significant conclusions can be made regarding the contemporaneous causal structure between various countries and regions. These findings can provide valuable insights into their intertwined economic and financial relationships. We present our main results in what follows.

From an integrated perspective, the graph presents a sparse structure, as evidenced by the small average number of edges of each node. To be precise, each node is connected with an average of merely 3.03 other nodes. This observation underscores the selective nature of interactions and relationships between the represented entities. In terms of edge number, three countries including the United Kingdom, the United States, and Germany stand out for their numbers of edges to other countries (or regions). This is consistent with the findings in \citet{giudici2016Graphical}.   
 
For the nature of the edges, let consider the in-degree and out-degree of nodes (representing countries or regions). The in-degree of a node refers to the number of edges directed towards it, signifying the extent to which a country is a receiver of banking connections from other countries or regions. Notably, the United Kingdom and the  United States exhibit high in-degrees, with values of 11 and 17 respectively. It indicates that many other countries (or regions) are directly connected to the United Kingdom and the  United States.  Conversely, the out-degree of a node  represents the number of edges going out from it. In this regard, the United Kingdom emerges as the country with the highest out-degree, followed by the  United States. This indicates that both the United Kingdom and the  United States are not only major receivers but also significant providers of banking services at an international level. The dual role as both a key recipient and provider underscores the integral position of the United Kingdom and the United States within the cross-border banking network.

Next, we analyze the relationships among these 26 countries (or regions) based on their economic classifications. Based on \citet{wooldridge2020implications}, we know advanced economies include Belgium, Canada, Switzerland, Germany, Denmark, Spain, France, the United Kingdom, Italy, Japan, Netherlands, Norway, Sweden, and the United States. Emerging market economies include China, Hong Kong SAR, India, South Korea, and Singapore. The off-shore markets include Cayman Islands, Bahamas, Hong Kong SAR, and Singapore, while Eurozone countries include Germany, France, Italy, Netherlands, Belgium, Luxembourg, Ireland, Spain, Austria, and Finland. For the edges across different economic classifications, a distinction becomes evident between emerging market economies and advanced economies. As the graph illustrates, emerging market economies tend to have fewer nodal degrees than advanced economies, which emphasizes the developmental disparities between the two categories, as also noted in \citet{wooldridge2020implications}.

For the edges within each classification, China is most connected within the emerging market economies. Particularly, China has a positive influence on Hong Kong SAR in the domain of cross-border banking liabilities. This is understandable given mainland China's pivotal role in bolstering Hong Kong's economic progress. The strong commitment of the Chinese government to Hong Kong, as highlighted by \citet{zhou2019How}, further supports this.
A distinct feature within the Eurozone's representation is the marked absence of linkages among its constituent nations (or regions), except for Germany. This can be explained by Germany's central role in the Eurozone, a view supported by \citet{bartlett2016interdependence}.

\section{Conclusion}\label{sec:summary}
In this paper, we introduce a novel model averaging approach designed to handle the uncertainties associated with node connection relationships in DAGs. This method involves generating a series of nested candidate models and aggregating them by assigning weights that are optimized by minimizing a penalized negative log-likelihood criterion. The theoretical findings presented herein suggest that our proposed method is capable of attaining asymptotic optimality as well as parameter consistency. The simulation studies confirm that our method provides high levels of prediction and estimation accuracy across diverse scenarios.


For future research, we will extend our results to more general noise assumptions, such as unequal variances or non-Gaussian errors.
Additionally, we plan to develop model averaging techniques for alternative graph structures, including complete partially directed acyclic graphs and dynamic graphs.

\bibliography{bibliography.bib}

@article{zouOptimalModelAveraging2022a,
  title = {Optimal Model Averaging for Divergent-Dimensional Poisson Regressions},
  author = {Zou, Jiahui and Wang, Wendun and Zhang, Xinyu and Zou, Guohua},
  year = {2022}, 
  journal = {Econometric Reviews},
  volume = {41},
  number = {7},
  pages = {775--805} 
}

@article{liu2023Frequentist,
  title = {Frequentist Model Averaging for Undirected {Gaussian} Graphical Models},
  author = {Liu, Huihang and Zhang, Xinyu},
  year = {2023}, 
  journal = {Biometrics},
  volume = {79},
  number = {3},
  pages = {2050--2062} 
}

@article{TONZER2015JFS,
title = {Cross-border interbank networks, banking risk and contagion},
journal = {Journal of Financial Stability},
volume = {18},
pages = {19-32},
year = {2015}, 
author = {Lena Tonzer} 
}

@article{Bremus2015JBF,
title = {Cross-border banking, bank market structures and market power: Theory and cross-country evidence},
journal = {Journal of Banking \& Finance},
volume = {50},
pages = {242-259},
year = {2015}, 
author = {Franziska M. Bremus}  
}

@article{yuan2019constrained,
  author = {Yuan, Yiping and Shen, Xiaotong and Pan, Wei and Wang, Zizhuo},
  title = {Constrained likelihood for reconstructing a directed acyclic {Gaussian} graph},
  journal = {Biometrika},
  volume = {106},
  number = {1},
  pages = {109-125},
  year = {2019},
  month = {12} 
}

@article{kalisch2007estimating,
  title={Estimating high-dimensional directed acyclic graphs with the {PC}-algorithm.},
  author={Kalisch, Markus and B{\"u}hlman, Peter},
  journal={Journal of Machine Learning Research},
  volume={8},
  number={3},
  year={2007},
  pages={613--636}
}

@article{hansen2007least,
  title={Least squares model averaging},
  author={Hansen, Bruce E.},
  journal={Econometrica},
  volume={75},
  number={4},
  pages={1175--1189},
  year={2007},
  publisher={Wiley Online Library}
}

@article{chickering2002optimal,
 title={Optimal structure identification with greedy search},
  author={Chickering, David Maxwell},
  journal={Journal of Machine Learning Research},
  volume={3},
  pages={507--554},
  year={2002}
}

@book{korb2010bayesian,
  author       = {Kevin B. Korb and Ann E. Nicholson},
  title        = {Bayesian Artificial Intelligence},
  edition      = {2nd},
  publisher    = {Chemical Rubber Company Press},
  address      = {Florida, USA},
  year         = {2010},
  series       = {Computer Science \& Data Analysis},
  serieseditor = {John Lafferty and David Madigan and Fionn Murtagh and Padhraic Smyth},
  isbn         = {978-1-58488-387-1}
}

@article{zhang2019inference,
  title={Inference after model averaging in linear regression models},
  author={Zhang, Xinyu and Liu, Chu-An},
  journal={Econometric Theory},
  volume={35},
  number={4},
  pages={816--841},
  year={2019},
  publisher={Cambridge University Press}
}

@article{meinshausen2006high,
author = {Meinshausen, Nicolai and B{\"{u}}hlmann, Peter}, 
journal = {The Annals of Statistics},
number = {3},
pages = {1436--1462},
title = {High-dimensional graphs and variable selection with the Lasso},
volume = {34},
year = {2006}
}

@article{zhang2016optimal,
author = {Zhang, Xinyu and Yu, Dalei and Zou, Guohua and Liang, Hua}, 
journal = {Journal of the American Statistical Association}, 
number = {516},
pages = {1775--1790},
title = {Optimal Model Averaging Estimation for Generalized Linear Models and Generalized Linear Mixed-Effects Models},
volume = {111},
year = {2016}
}

@article{scutari2010bnlearn,
author = {Scutari, Marco}, 
journal = {Journal of Statistical Software}, 
number = {3},
title = {Learning Bayesian networks with the {bnlearn} {R} Package},
volume = {35},
year = {2010},
pages={1--22}
}

@book{nagarajan2013bayesian,
address = {New York, USA},
author = {Nagarajan, Radhakrishnan and Scutari, Marco and L{\`{e}}bre, Sophie}, 
publisher = {Springer},
title = {Bayesian Networks in R},
year = {2013}
}

@article{peters2014identifiability,
author = {Peters, J. and B{\"{u}}hlmann, P.},
journal = {Biometrika},
number = {1},
pages = {219--228},
publisher = {[Oxford University Press, Biometrika Trust]},
title = {Identifiability of {Gaussian} structural equation models with equal error variances},
volume = {101},
year = {2014} 
}

@article{suter2021BiDAG,
 title={Bayesian Structure Learning and Sampling of {Bayesian} Networks with the {R} Package BiDAG},
 volume={105}, 
 number={9},
 journal={Journal of Statistical Software},
 author={Suter, Polina and Kuipers, Jack and Moffa, Giusi and Beerenwinkel, Niko},
 year={2023},
 pages={1--31}
}

@book{johnson2001handbook,
  title       = {Handbook of the Geometry of Banach Spaces},
  editor      = {Johnson, W. B. and Lindenstrauss, Joram},
  publisher   = {Elsevier / North Holland},
  address     = {Amsterdam},
  year        = {2001}
}

@article{imbens2020Potential,
  title = {Potential Outcome and Directed Acyclic Graph Approaches to Causality: Relevance for Empirical Practice in Economics},
  shorttitle = {Potential Outcome and Directed Acyclic Graph Approaches to Causality},
  author = {Imbens, Guido W.},
  year = {2020},
  month = dec,
  journal = {Journal of Economic Literature},
  volume = {58},
  number = {4},
  pages = {1129--1179}
}

@article{giudici2016Graphical,
  title = {Graphical Network Models for International Financial Flows},
  author = {Giudici, P. and Spelta, A.},
  year = {2016}, 
  journal = {Journal of Business \& Economic Statistics},
  volume = {34},
  number = {1},
  pages = {128--138} 
}

@misc{wooldridge2020implications,
  title={Implications of financial market development for financial stability in emerging market economies},
  author={Wooldridge, Philip},
  howpublished = {Bank for International Settlements},
  year={2020} 
}

@article{bartlett2016interdependence,
  author      = {William Bartlett and Ivana Prica},
  title       = {Interdependence between Core and Peripheries of the European Economy: Secular Stagnation and Growth in the Western Balkans},
  journal     = {LSE Europe in Question Discussion Paper Series},  
  year        = {2016},
  note        = {Paper No.\ 104, London School of Economics}
}

@misc{zhou2019How,
  title = {{How Hong Kong survived the 1998 financial crisis}},
  author = {Zhou, Minxi},
  year = {2019},
  month = aug,
  day = {14},
  howpublished = {CGTN},
  note={{https://news.cgtn.com/news/2019-08-14/How-Hong-Kong-survived-the-1998-financial-crisis-J9lwvZrsNq/index.html}}
}

@article{shojaie2010Penalized,
title = {Penalized Likelihood Methods for Estimation of Sparse High-Dimensional Directed Acyclic Graphs},
author = {Shojaie, A. and Michailidis, G.},
year = {2010},
month = sep,
journal = {Biometrika},
volume = {97},
number = {3},
pages = {519--538}
}

@article{fu2013Learning,
title = {Learning Sparse Causal {Gaussian} Networks With Experimental Intervention: Regularization and Coordinate Descent},
author = {Fu, Fei and Zhou, Qing},
year = {2013}, 
journal = {Journal of the American Statistical Association},
volume = {108},
number = {501},
pages = {288--300}
}

@incollection{pearl2012causal,
title={The causal foundations of structural equation modeling},
booktitle = {Handbook of Structural Equation Modeling},
editor = {R. H. Hoyle},
publisher = {Guilford Press},
author={Pearl, Judea},
year={2012},
pages = {68--91},
address={New York, USA}
}

@article{kaplan2016Bayesian,
title = {Bayesian Model Averaging Over Directed Acyclic Graphs With Implications for the Predictive Performance of Structural Equation Models},
author = {Kaplan, David and Lee, Chansoon},
year = {2016}, 
journal = {Structural Equation Modeling: A Multidisciplinary Journal},
volume = {23},
number = {3},
pages = {343--353}
}

@article{zhang2020parsimonious,
author = {Zhang, Xinyu and Zou, Guohua and Liang, Hua and Carroll, Raymond J.},
journal = {Journal of the American Statistical Association},
number = {530},
pages = {972--984},
publisher = {Taylor {\&} Francis},
title = {Parsimonious model averaging with a diverging number of parameters},
volume = {115},
year = {2020}
}

@article{Lehrer2022MS,
author = {Lehrer, Steven F. and Xie, Tian},
title = {The Bigger Picture: Combining Econometrics with Analytics Improves Forecasts of Movie Success},
journal = {Management Science},
volume = {68},
number = {1},
pages = {189-210},
year = {2022}
}

@article{Liang2011JASA,
author = {Liang, Hua and Zou, Guohua and Wan, Alan T.K.  and Zhang, Xinyu},
title = {Optimal Weight Choice for Frequentist Model Average Estimators},
journal = {Journal of the American Statistical Association},
volume = {106},
number = {495},
pages = {1053-1066},
year = {2011},
publisher = {Taylor & Francis} 
}

@inproceedings{ghoshal2018learning,
  title = {Learning Linear Structural Equation Models in Polynomial Time and Sample Complexity},
  booktitle = {Proceedings of the {{Twenty-First International Conference}} on {{Artificial Intelligence}} and {{Statistics}}},
  author = {Ghoshal, Asish and Honorio, Jean},
  year = {2018},
  pages = {1466--1475},
  publisher = {PMLR},
  issn = {2640-3498},
  address={Lanzarote, Spain}
}

@article{park2020identifiability,
  title = {Identifiability of Additive Noise Models Using Conditional Variances},
  author = {Park, Gunwoong},
  year = {2020},
  journal = {Journal of Machine Learning Research},
  volume = {21},
  number = {75},
  pages = {1--34},
  issn = {1533-7928},
  urldate = {2024-10-26}
}

@article{li2020likelihood,
author = {Chunlin Li, Xiaotong Shen and Wei Pan},
title = {Likelihood Ratio Tests for a Large Directed Acyclic Graph},
journal = {Journal of the American Statistical Association},
volume = {115},
number = {531},
pages = {1304--1319},
year = {2020}

}

@article{hansen2014QE,
  title={Model averaging, asymptotic risk, and regressor groups},
  author={Hansen, Bruce E},
  journal={Quantitative Economics},
  volume={5},
  number={3},
  pages={495--530},
  year={2014},
  publisher={Wiley Online Library}
}

@article{cheng2019QE,
  title={On uniform asymptotic risk of averaging GMM estimators},
  author={Cheng, Xu and Liao, Zhipeng and Shi, Ruoyao},
  journal={Quantitative Economics},
  volume={10},
  number={3},
  pages={931--979},
  year={2019},
  publisher={Wiley Online Library}
}

@article{zhang2011focused,
  title = {Focused Information Criterion and Model Averaging for Generalized Additive Partial Linear Models},
  author = {Zhang, Xinyu and Liang, Hua},
  year = {2011},
  journal = {Annals of Statistics},
  volume = {39},
  number = {1},
  pages = {174--200},
  issn = {00905364},
  doi = {10.1214/10-AOS832},
}

@article{heckerman1995learning,
  title = {Learning Bayesian Networks: The Combination of Knowledge and Statistical Data},
  shorttitle = {Learning {{Bayesian Networks}}},
  author = {Heckerman, David and Geiger, Dan and Chickering, David M.},
  year = {1995},
  month = sep,
  journal = {Machine Learning},
  volume = {20},
  number = {3},
  pages = {197--243},
  issn = {1573-0565},
  doi = {10.1023/A:1022623210503},
}

@article{tsamardinos2006maxmin,
  title = {The Max-Min Hill-Climbing {Bayesian} Network Structure Learning Algorithm},
  author = {Tsamardinos, Ioannis and Brown, Laura E. and Aliferis, Constantin F.},
  year = {2006},
  month = oct,
  journal = {Machine Learning},
  volume = {65},
  number = {1},
  pages = {31--78},
  issn = {1573-0565},
  doi = {10.1007/s10994-006-6889-7},
}

@article{heckman2010building,
  title = {Building Bridges between Structural and Program Evaluation Approaches to Evaluating Policy},
  author = {Heckman, James J.},
  year = {2010},
  month = jun,
  journal = {Journal of Economic Literature},
  volume = {48},
  number = {2},
  pages = {356--398},
  issn = {0022-0515},
  doi = {10.1257/jel.48.2.356},
}

@article{durlauf2005growth,
  title={Growth econometrics},
  author={Durlauf, Steven N and Johnson, Paul A and Temple, Jonathan RW},
  journal={Handbook of economic growth},
  volume={1},
  pages={555--677},
  year={2005},
  publisher={Elsevier}
}

@article{leamer1983lets,
  title = {Let's Take the Con Out of Econometrics},
  author = {Leamer, Edward E.},
  year = {1983},
  journal = {The American Economic Review},
  volume = {73},
  number = {1},
  eprint = {1803924},
  eprinttype = {jstor},
  pages = {31--43},
  publisher = {American Economic Association},
  issn = {0002-8282},
  urldate = {2025-07-04}
}

@incollection{card1999chapter,
  title = {The Causal Effect of Education on Earnings},
  booktitle = {Handbook of {{Labor Economics}}},
  author = {Card, David},
  editor = {Ashenfelter, Orley C. and Card, David},
  year = {1999},
  month = jan,
  volume = {3},
  pages = {1801--1863},
  publisher = {Elsevier},
  doi = {10.1016/S1573-4463(99)03011-4},
}

@book{gruber2016public,
  title = {Public Finance and Public Policy},
  author = {Gruber, Jonathan},
  year = {2016},
 edition   = {5th},  
  publisher = {Worth Publishers},
  address = {New York},
  isbn = {978-1-319-15416-5},
  langid = {english},
  lccn = {HJ141 G88}
}

@article{Han2016JASA,
author = {Sung Won Han and Gong Chen and Myun-Seok Cheon and Hua Zhong},
title = {Estimation of Directed Acyclic Graphs Through Two-Stage Adaptive Lasso for Gene Network Inference},
journal = {Journal of the American Statistical Association},
volume = {111},
number = {515},
pages = {1004--1019},
year = {2016},
publisher = {Taylor \& Francis},
doi = {10.1080/01621459.2016.1142880},
URL = {https://doi.org/10.1080/01621459.2016.1142880},
eprint = {https://doi.org/10.1080/01621459.2016.1142880}

}

@article{ellis2008jasa,
  author = {Byron Ellis and Wing Hung Wong},
  title   = {Learning Causal {B}ayesian Network Structures from Experimental Data},
  journal = {Journal of the American Statistical Association},
  year    = {2008},
  volume  = {103},
  number  = {482},
  pages   = {778--789},
 URL = {http://www.jstor.org/stable/27640100}
}

@article{vanderweele2010jrssb,
  author = {Tyler J. VanderWeele and James M. Robins},
  title   = {Signed Directed Acyclic Graphs for Causal Inference},
  journal = {Journal of the Royal Statistical Society: Series B (Statistical Methodology)},
  year    = {2010},
  volume  = {72},
  number  = {1},
  pages   = {111--127},
  URL = {http://www.jstor.org/stable/40541577}
}

\end{document}